\documentclass[reprint,tightenlines,prx,superscriptaddress]{revtex4-2}

\usepackage{graphicx}
\usepackage{xcolor}
\usepackage{amsmath}
\usepackage{amsfonts}
\usepackage{amssymb}
\usepackage{amsthm}
\usepackage{fullpage}
\usepackage{braket}
\usepackage{tikz}
\usepackage{quantikz}
\usepackage{adjustbox}
\usepackage{tabularx}
\usepackage{floatrow}
\usepackage{booktabs}

\usepackage[caption=false, label font=bf]{subfig}

\renewcommand\bra[1]{{\langle{#1}|}}
\renewcommand\ket[1]{{|{#1}\rangle}}

 {\begin{list}{}%
         {\setlength{\leftmargin}{#1}}%
         \item[]%
 }
 {\end{list}}
 
\newcolumntype{Y}{>{\centering\arraybackslash}X}
\newcolumntype{Z}{>{\raggedright\arraybackslash}X}

\begin{document}


\title{Prospects for NMR Spectral Prediction on Fault--Tolerant Quantum Computers}

\author{J. E. Elenewski}
\email[]{justin.elenewski@ll.mit.edu}
\affiliation{MIT Lincoln Laboratory, Lexington, Massachusetts 02421, USA}
\author{C. M. Camara}
\affiliation{Department of Pediatric Oncology and the Linde Program in Cancer Chemical Biology, Dana--Farber Cancer Institute, Boston, MA, USA}
\author{A. Kalev}
\email[]{amirk@isi.edu}
\affiliation{Information Sciences Institute, University of Southern California, Arlington, VA 22203, USA}
\affiliation{Department of Physics and Astronomy, and Center for Quantum Information Science \& Technology, University of Southern California, Los Angeles, California 90089, USA}

\begin{abstract} 
Advanced atomic magnetometers have made it possible to acquire nuclear magnetic resonance spectra  in zero to ultralow magnetic fields.  This regime carries the benefit of compact, low--cost instrumentation with reduced spin relaxation effects and the ability to probe phenomena  that are inaccessible in conventional high--field experiments.  A tradeoff is that the resulting spectra must  be interpreted using simulations that are taxing for classical computation.  Working by example for  small--molecule and protein spectroscopy, we demonstrate that these simulations are a promising target for fault--tolerant quantum computation. Our holistic analysis spans from input selection to the construction of explicit circuits for qubitized quantum dynamics. By maintaining parity with experimental requirements, we demonstrate how certain cases might be especially promising for early fault--tolerant architectures.  Notably, valuable natural products and small proteins could be simulated with a few hundred logical qubits and less than $10^{12}$ $T$--gates, corresponding to days of runtime with foreseeable hardware based on spin or superconducting qubits.
\end{abstract}
\maketitle


\section{Introduction}
\par Nuclear magnetic resonance spectroscopy (NMR) has a notable ability to resolve the atomic--scale structures and dynamics of analytes across a  hierarchy of timescales  \cite{Ernst1990}.  This  breadth has  ensconced NMR as an analytical gold standard in chemistry, physics, and materials science.  While most use cases are related to structure determination, more nuanced studies of reaction dynamics and macromolecular function have had an outsized impact \cite{ Hu2021, Marusic2023, Fontana2023}. In particular, the ability to probe  near--native environments has delivered critical  insights into the folding and interactions of biomolecules such as  proteins and nucleic acids \cite{Rule2006,Sprangers2007,Weingarth2013,Rosenzweig2016,Puthenveetil2019,Bayer2020,Reif2021,Hu2021}.  On a related note, magnetic resonance imaging and spatially--resolved NMR spectroscopy have become standard diagnostics in clinical medicine \cite{Cecil2013,Oz2014,Tognarelli2015, Brown2014,Soares2016,Westbrook2018}.

\par The development of NMR has generally pursued higher spectral resolution over other features. This has led to instruments with  large static magnetic fields, $B_0$, which give a broad frequency dispersion $\Delta \omega \sim B_0$ among resonances and increase the fraction $P \sim B_0 / k_BT$ of polarized spins \cite{Ernst1990,Rule2006}.  While the resulting spectra are less crowded and exhibit better  signal--to--noise characteristics, the associated hardware is also costly and has high operational overhead.  Nonethless, advances  in atomic magnetometry have recently opened a path toward NMR spectroscopy in zero--to--ultralow fields (ZULF; below 100 nT) where  hyperpolarization  can deliver  spin populations  that surpass high--field experiments by several orders of magnitude \cite{Bernarding2006, Blanchard2021, DeVience2021, Theis2011} (reviewed in Ref.~\cite{Barskiy2024}). This is of technological interest since low--field instruments are comparatively inexpensive, portable, and cryogen--free.   ZULF experiments have also been shown to mitigate field--associated relaxation and can target an insightful regime for spin physics   \cite{Blanchard2013, Emondts2014, Blanchard2015,Appelt2010}.   However, only the simplest of these spectra can be interpreted without simulation.

\begin{figure}
\begin{center}
\includegraphics[width=0.75\columnwidth]{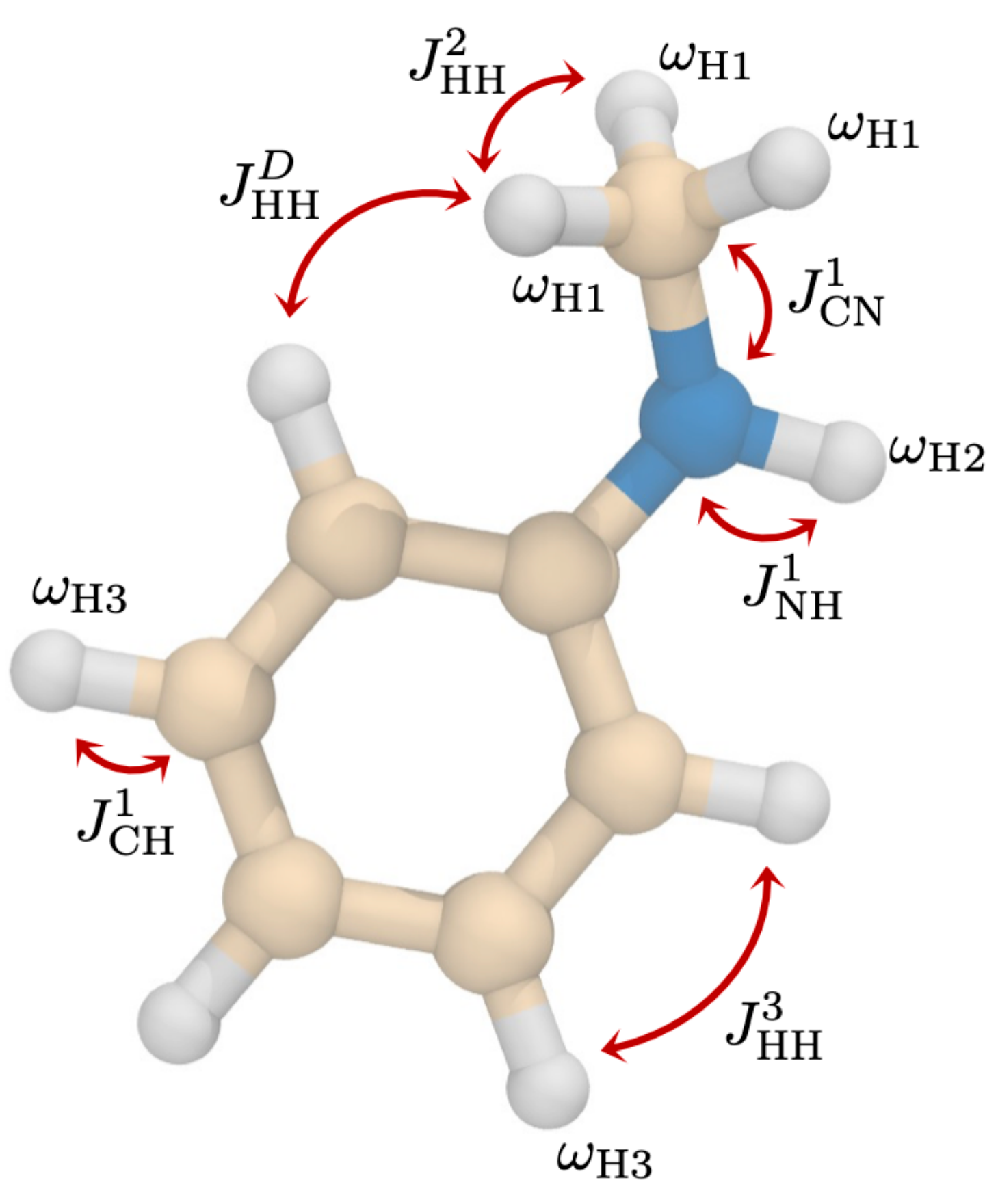}
\end{center}
\caption{{\bf Components of a nuclear spin Hamiltonian.} Representative chemical shifts and couplings, as illustrated using the organic molecule $N$-methylaniline. Chemically equivalent protons share the same chemical shift, which are labeled for $\omega_{\text{H}1}$, $\omega_{\text{H}2}$, and  $\omega_{\text{H}3}$. Electronically--mediated $J$--couplings are shown for select homonuclear ($J^n_\text{HH}$) and heteronuclear cases ($J^n_\text{CH}$, $J^n_\text{NH}$). These differ in the number of intermediate bonds $n$ between coupled spins. A dipolar coupling $J^D_\text{HH}$ is also shown. Dipolar couplings are defined between atoms independent of the  $J$--coupling.}
\label{fig:shifts_and_couplings}
\end{figure}

\par The ultralow--field regime is characterized by long--range vector spin couplings with  little separation between energy scales \cite{Karabanov2011}.  These features are resource--intensive for classical algorithms but are particularly  natural for their quantum counterparts \cite{Wilzewski2017,Seetharam2023}. Attesting to this, several NMR--related tasks have been proposed for noisy intermediate--scale quantum (NISQ) computers \cite{Algaba2022,OBrien2022, Seetharam2023, Burov2024}.  This is perhaps unsurprising, since the calculation of NMR spectra can be directly cast as an exercise in quantum simulation.

\par  In this contribution, we go beyond prior work to consider  NMR spectral prediction in the context of fault--tolerant quantum computation (FTQC). Our discussion is framed around experiments in the ZULF regime, though similar considerations hold for strongly--coupled spin systems and conventional solid--state NMR experiments at high field.  We  begin by identifying realistic parameters and targets for computation, including the specification of input molecules across a range of application scales and complexities. Guided by these data, we analyze circuit level implementations for state--of--the art qubitized algorithms as well as optimized problem encodings that mitigate logical resource overhead. Finally, we estimate the physical requirements for select calculations when using a conventional surface code architecture and a lattice surgery scheme based on sequential Pauli product rotations \cite{Litinski2019}.   We predict that meaningful spectra might be computed with  overhead comparable to or less than other guidestar problems, such as factoring 2048--bit integers using Shor's algorithm \cite{Gidney2021} or simulating classically hard instances of the Fermi--Hubbard model and spin liquid Hamiltonians. This suggests that NMR spectral prediction is a robust computational objective and a potential application for early fault--tolerant quantum hardware.

\section{Simulation Physics}

\subsection{Experimental Context}

\par The connection between NMR and quantum computation becomes  apparent when we consider a simple experiment.  In this setting, nuclear spin evolution is initiated using a discrete series of radiofrequency (RF) pulses that are tuned to the Lamor frequency of a target nucleus.  This causes the spin system to precess and generates a magnetization profile that is captured by  pickup coils in the spectrometer. The resulting signal $\mathcal{S}(t) $ is called the free--induction decay (FID). As  the name suggests, this signal is collected as the system equilibrates with its environment.   Our interpretable spectrum  is a sparse, frequency--domain counterpart to this profile $\mathcal{S}(\omega) = \int_0^{t_\text{max}}  \mathcal{S}(t) \, e^{\imath\omega t} \, dt$, where $t_\text{max}$ is the FID  duration. We can simulate this by computing a spin--spin correlator, 
\begin{equation}\label{ref:nmr_time_domain}
\langle S_\text{tot}(t) S_\text{tot}(0) \rangle = \text{tr} \, [e^{\imath H t} S_\text{tot} e^{-\imath H t} S_\text{tot}  \rho],
\end{equation}
\noindent where $H$ is the system's Hamiltonian, $\rho$ is its density matrix, and the operator $S_\text{tot} = \sum_k S_k$ captures a total spin. While a typical spectrometer will use phase--sensitive quadrature to detect $S^+ =  S^x + \imath S^y$, this is tied to the experimental apparatus and  control pulses. The problem is simpler for computation,   where  we only need $S_\text{tot} = S^z_\text{tot} = \sum_k S^z_k$ to capture the location of resonances \footnote{The most straightforward digital quantum simulations will reproduce a pure state density matrix as opposed to the mixed state generated by relaxation operators.  Instead, we treat relaxation through the decaying exponential factor in Eq.~\ref{ref:nmr_frequency_domain}. Thus, the longitudinal and transverse magnetization profiles become equivalent for determining resonances in the NMR spectrum. It may be prudent to work with $S^+$ or $S^-$ when developing pulse sequences due to the difference in objectives and potential methodology.}.  This means that the simulated spectrum $\tilde{\mathcal{S}}(\omega) $ will take a simple form:
\begin{equation} \label{ref:nmr_frequency_domain}
\tilde{\mathcal{S}}(\omega) = \int_0^{t_\text{max}} \,   \langle S^z_\text{tot}(t) S^z_\text{tot}(0) \rangle \, e^{\imath\omega t - \gamma_2 t} \, dt.
\end{equation}

\noindent Note the factor containing $\gamma_2 = T_2^{-1}$, which captures relaxation mechanisms that lead to decoherence and dephasing.  This can be  handled using a relaxation superoperator in more nuanced treatments.

\subsection{Nuclear Spin Hamiltonian}

\par NMR spectroscopy is predicated on the  timescale disparity between electronic and nuclear dynamics in a material \cite{Ernst1990}. This effectively decouples the two subsystems, though the faster electronic component still mediates the nuclear spins  through time--averaged parameters. The canonical Hamiltonian for a system of $N$ nuclear spins follows accordingly:

\begin{equation}\label{eq:nmr_hamiltonian}
\begin{split}
H & =  -\hbar \sum_k \gamma_k (1 - \boldsymbol{\delta}_k) \, \mathbf{I}_k  \cdot \mathbf{B}_0  - \hbar \sum_k \gamma_k \mathbf{I}_k  \cdot \mathbf{B}_{RF} (t)\\
&    + \hbar \sum_{k \neq l} J_{kl} \, \mathbf{I}_k \cdot \mathbf{I}_l  \\
& + \sum_{k \neq l} b_{k l} \left[\mathbf{I}_k \cdot \mathbf{I}_l - 3 \frac{1}{\vert\vert \mathbf{r}_{kl} \vert\vert^2} (\mathbf{I}_k \cdot \mathbf{r}_{kl}) (\mathbf{I}_l \cdot \mathbf{r}_{kl}) \right].\\
\end{split}
\end{equation}

\noindent We adopt a  convention  where $\mathbf{I}_k = \mathbf{S}_k / \hbar = (S_k^x, S_k^y, S_k^z) / \hbar$  is a vector of spin angular momentum operators for the $k$-th spin. These are defined as $S^\alpha_k = \hbar \sigma^\alpha/2$ with $\sigma^\alpha$  a spin-1/2 Pauli operator.

\par The first term is a Zeeman coupling between the $k$-th atom's nuclear spin and a static, external magnetic field $\mathbf{B}_0$.    Both the nuclear gyromagnetic ratio $\gamma_k$ and the diamagnetic shielding tensor $\boldsymbol{\delta}_k$ appear here, reflecting a dependence on the nuclear species and its electronic environment. This confers a distinctive ``chemical shift'' frequency to each nucleus that vanishes at zero field. The absence of this feature causes different resonances to overlap, complicating the interpretation of ultralow field spectra \footnote{While there will be a chemical shift at small but finite fields, resonances can still be indistinguishable due to line broadening.}. The second term constitues a similar coupling to the magnetic component of a classical RF pulse. This is useful when simulating specific control pulses but is unnecessary when assigning  resonances.

\par The second and third terms are scalar and dipolar spin--spin couplings, respectively. These differ in terms of prominence and prevalence.  The scalar coupling (or $J$--coupling) is an effective hyperfine shift mediated by the electronic environment of chemically bonded atoms. This perturbs the stronger Zeeman contribution at high--field, splitting the resonance at a given chemical shift frequency. Conversely, the scalar coupling shifts resonances away from $\omega = 0$ in ultralow--field experiments and gives rise to the main spectral features.  Scalar couplings are a sensitive probe of the spin neighborhood and   molecular geometry (Fig.~\ref{fig:shifts_and_couplings}) \footnote{Stated generally, the scalar coupling will be a tensor quantity. However, only the scalar isotropic component is relevant for the liquid--phase experiments that we consider. Other protocols can access different components of the coupling tensor and extract different information.}.  The last term is a dipolar coupling  between  spatially--proximate spins, where $\mathbf{r}_{kl}$ is a vector between the $k$-th and $l$-th nuclei. For convenience, we have absorbed the spatial dependence into the dipolar coupling strength $b_{kl} = \mu_0 \gamma_k \gamma_l \hbar / 4\pi \vert\vert{\mathbf{r}_{kl}} \vert\vert^3$. These contributions average to zero in liquids yet are prominent in the solid--state or in weakly--aligning  media.  Both scalar and dipolar couplings appear similar in spectra.

\par These assumptions reduce the ZULF Hamiltonian to an effective Heisenberg Hamiltonian:
\begin{equation}
H = \sum_{k \neq l} \tilde{J}_{kl} \, \vec{\sigma}_k \cdot \vec{\sigma}_l + \sum_{\substack{k \neq l \\ \alpha, \beta}} \tilde{D}_{kl}^{\alpha\beta} \, \sigma_k^\alpha \sigma_l^\beta  - \sum_k \vec{h}_k \cdot \vec{\sigma}_k. 
\end{equation}

\noindent This will be the target of our analysis. Here, the rescaled scalar $\tilde{J}_{kl}$ and dipolar coupling components $\tilde{D}_{kl}^{\alpha\beta}$ are specified for a fixed atomic configuration with $\alpha,\beta \in \{x,y,z\}$.  We have included the field--dependent Zeeman terms for completeness even though they vanish in our calculations. The structure of this Hamiltonian makes the classical hardness of ZULF simulations apparent \cite{Appelt2010}. While certain maneuvers can simplify high--field calculations --- including perturbative strategies and reduced fidelity  targets due to relaxation --- these are inapplicable at ultralow fields \cite{Emondts2014,Tayler2019b}. Furthermore, the conjunction of low noise / weak relaxation and long--range interactions is especially problematic for classical tensor network simulations \cite{Zhou2020,Ayral2023}.


\subsection{Input Specification} \label{sec:hamiltonian_inputs}

\par  We generate resource estimates for a large pool of molecular structures.  These inputs are dawn from datasets that broadly sample chemical parameter space, allowing us to capture overhead for real--world problems.  In order of increasing complexity they include (i) fragment and screening databases for drug discovery, (ii) marketed small--molecule pharmaceuticals, (iii) biological secondary metabolites or ``natural products'' and (iv) a series of representative biomolecules (proteins).  Molecular structures are mapped to nuclear spin Hamiltonians by identifying plausible couplings between  spin--1/2 nuclei. We write $J_{\text{AB}}^k$ to denote scalar ($J$) couplings between nuclei of species A, B that are separated by $k$ bonds and use $J_{\text{AB}}^D$ to denote dipolar couplings between the same nuclei \footnote{ We drop the isotopic label in subscripts since all nuclei are assumed to be the most abundant spin-1/2 isotope.}.  Our large datasets make it infeasible to use experimental coupling parameters for each molecule. Instead, we defer to representative scalar couplings for each chemical environment  and calculate dipolar couplings based on molecular geometry \cite{Kemp1986, Hogben2011}.   Note that electronic structure calculations can be used to estimate spin couplings when  reference data are unavailable \cite{Cremer2007, Helgaker2016}.

\par We consider several  schemes based on the coupled nuclei that are detected. The simplest case (denoted {\bf Proton})  is based on homonuclear scalar couplings $J_{\text{HH}}^k$ between protons ($^{1}$H) that are separated by up to four bonds.  This is common for many high--field NMR spectra and has recently become feasible at ultralow fields \cite{DeVience2021,DeVience2022}. We also consider a case with both  heteronuclear and homonuclear couplings ({\bf Heteronuclear}) across $^{1}$H, $^{13}$C, and $^{15}$N nuclei. This resembles many ZULF experiments to date, which utilize at least one isotopically--labeled heteronucleus. We assume that all of the respective atoms in a given molecule will be one of these magnetically active isotopes, though this is only true for a purposefully enriched sample. Thus, our estimates will be an upper bound on simulation complexity for many experiments. Note that these schemes are based on scalar couplings, which can fragment the spin network into disconnected clusters.  The required qubit count will be proportional to the number of spins in the largest cluster, capping  spatial overhead and aiding parallel simulation on a large quantum computer.  We  assume that sufficiently small clusters can  be classically outsourced.

\par Dipolar interactions can become  observable in solid--state samples or when using alignment media.  When relevant, we accommodate the former ({\bf + Dipolar}) by including dipolar terms of strength $b_{kl}$ between $S = 1/2$ nuclei.    The case of weak alignment is  similar, though  we rescale $b_{kl} \mapsto \kappa \, b_{kl}$ with $\kappa = 1.0 \times 10^{-3}$ to emulate the weaker residual dipolar couplings ({\bf + RDCs}) \cite{Tzvetkova2019}.  These are helpful when resolving molecular orientation and have been detected in ZULF experiments \cite{Blanchard2015}. We only include dipolar couplings between nuclei separated by $\vert \vert \mathbf{r}_{kl} \vert \vert \leq 4.0$\,\AA\, which is reasonable for RDC experiments.  While this constitutes a sharp truncation for solid--state experiments, it allows us to compare overhead on an even footing.


\subsection{Simulation Parameters}

\par Our aim is to calculate $\langle S_\text{tot}^z(t) S_\text{tot}^z (0) \rangle$ at a series of timepoints up to a maximal sampling duration $t_\text{max}$.  This is done with a finite spectral resolution $\Delta \omega$ in the frequency domain that is fixed by the experimental spin--spin relaxation time $\Delta \omega \propto 1/T_2$. Relaxation also sets a minimum fidelity target $F(t)$  each  evolution, with fidelity decreasing as $t$ approaches $T_2$.  In the simplest treatment, a dephasing rate of $\gamma = 1/T_2$ will cause each spin to independently decohere as $F \sim \exp[-\gamma t]$.  This  has been used to set error thresholds  of $\epsilon(t) \leq  (1-\exp[-N t/T_2])$ for NISQ algorithms, giving a  reduction in resource overhead \cite{Seetharam2023}. The utility of this is weaker for qubitized algorithms,  where overhead  scales logarithmically with precision and a high fidelity is needed to minimize algorithm repetitions.  As a compromise, we set a target of $\epsilon = (1-\exp[-t/T_2])$ for the target time evolution unitary but require that $ \epsilon \leq \epsilon_\text{max} = 5 \times 10^{-3}$.  Note that this limited--fidelity simulation is matched to experimental precision.  Thus, it fixes a baseline for other sources of error.

\par The maximal simulation time $t_\text{max}$ and sampling schedule are also critical parameters since they define  high-- and low--frequency cutoffs for $\mathcal{S}(\omega)$, respectively. With regard to $t_\text{max}$, experiments generally collect beyond $3T_2$ to minimize artifacts and often exceed  $t_\text{max} = 5T_2$  for a better signal--to--noise ratio  \footnote{The experimental budget is also guided by other parameters,  $t_\text{max} = n_\text{points} / 2W$, such as the desired spectral width $W$ and the number of points $n_\text{points}$ sampled in the signal. This budget is also guided by a tradeoff between satisfying Nyquist sampling requirements and practical spectrometer availability. Thus, it difficult to map literature parameters back to quantum simulations.  }.  However, fault--tolerant simulations are relaxation--free so our only concern is capturing experimentally detectable dynamics. We do not strive to rival experimental data but instead  assign  resonances when classical simulations are costly. To this end, we assume that most coherence pathways  have developed by $t_\text{max} = T_2$, with the remainder falling below the detection threshold.  A generic $T_2$ can be tricky to specify since ZULF relaxation can be an order of magnitude slower than the high-field setting. We  adopt a representative $t_\text{max} = T_2 = 1$ s as a compromise across  molecular scales.


\section{Quantum Algorithms}

\par Qubitized methods, such as  quantum signal processing (QSP) and its descendants, are among the most resource--efficient algorithms for fault--tolerant quantum simulation \cite{Low2017, Low2019, Gilyen2019, Martyn2021}. QSP itself can approximate $\mathcal{O}(t)$ a time--evolution unitary $\mathcal{U}(t)$ up to a precision $||\mathcal{O}(t) - \mathcal{U}(t)\vert\vert_2  \leq \epsilon$ using  $O(t + \log (1/\epsilon))$ queries to the block encoding of a time--independent Hamiltonian.  Spatial overhead is also mild since the encoding only requires $O(\log N)$ ancilla.  We briefly review these algorithms to frame our analysis.


\subsection{Block Encoding Strategy}

\par A prerequisite  for qubitized dynamics is the ability to represent our nonunitary Hamiltonian  $H$ in a quantum circuit. This is accomplished by block encoding $H$ in a larger unitary operator $U_H$,
\begin{equation}
U_H = \begin{pmatrix} H & \ast \\ \ast & \ast \end{pmatrix},
\end{equation}

\noindent which acts on a dilated Hilbert space.  By construction, we access $H$ using an ancillary selection state $\ket{G} \equiv \ket{0^m}$ so that,
\begin{equation}
H = \big(\bra{G}\otimes I_n\big) \, U_H \, \big(\ket{G} \otimes I_n\big),
\end{equation}

\noindent We are only concerned with the block that encodes $H$ and can leave the rest undefined.  The term qubitization alludes to this structure, where access for $H$ resembles a two--level system in terms of $\ket{G}$ and its orthogonal compliment. Note that our embedding must be a contraction, e.g., $\vert\vert H\vert\vert_2 \leq 1$, in order to maintain unitarity.   This typically requires the Hamiltonian to be rescaled $H \mapsto  H/\alpha$  and  can have a nontrivial impact on algorithm overhead.

\par Our nuclear spin Hamiltonian is characterized by Heisenberg couplings with a complicated topology.  We must map the associated spin network onto a one--dimensional  register of $N$ system qubits. Fortunately,  this reduces to simple indexing for $S = 1/2$. Each  term in $H = \sum_{i=1}^M c_i \Lambda_i$ corresponds to a Pauli string with two non--identity factors,
\begin{equation}
\Lambda_i = I^{\otimes p} \otimes P_k \otimes I^{\otimes q} \otimes P_l \otimes I^{\otimes (N-p-q-2)}.
\end{equation}

\noindent These terms are unitary meaning that $H$ is manifestly a linear combination of unitary operators (LCU). This allows us to block encode using a conventional strategy. 

\par To construct this encoding, we define a \emph{Select oracle} that acts on the system and selection registers,
\begin{equation}
U_\text{sel} = \sum_i \ket{i}\bra{i} \otimes \Lambda_i,
\end{equation}


\begin{figure*}
\begin{center}
\renewcommand\thesubfigure{\alph{subfigure}}
    \setlength{\labelsep}{2mm}
\sidesubfloat[] {
\begin{adjustbox}{scale=0.8}
\begin{quantikz}   
\lstick{\text{Phase} \;\;\;\; $\ket{0}$}      \qw & \qw        & \gate{R(\lambda_0,\phi_0,\theta_0)}  &  \octrl{1}                &   \gate{R(0,\phi_1,\theta_1)}          & \octrl{1}                      &  \gate{R(0,\phi_2,\theta_2)}      & \qw         & \dots \,\,\, & \octrl{1}                    &  \gate{R(0,\phi_{2d},\theta_{2d})} &\qw \\
\lstick{\text{Selection} \;\; $\ket{0^m}$}    \qw & \qw \qwbundle{m} & \qw                                  &  \gate[wires=2]{\mathcal{W}_H}      &   \qw                                  & \gate[wires=2]{\mathcal{W}_H^\dagger}    &  \qw                              & \qw         & \dots \,\,\, & \gate[wires=2]{\mathcal{W}_H}  & \qw &\qw                                \\
\lstick{Target \;\;\;\; $\ket{\psi}$}         \qw & \qw \qwbundle{n} & \qw                                  &                           &   \qw                                  &                                &  \qw                              & \qw         & \dots \,\,\, &                              & \qw &\qw                                    \end{quantikz}\end{adjustbox}
} \\ \vspace{2\baselineskip}
\sidesubfloat[] {
\begin{adjustbox}{scale=0.8}
\begin{quantikz}
\lstick{\text{Phase} \;\;\;}     \qw & \qw              &  \qw     & \octrl{1}            & \gate{Z}   & \qw  \\
\lstick{\text{Selection} \;\; }  \qw & \qw \qwbundle{m} &  \qw     & \gate[wires=2]{U_H}  & \octrl{-1} & \qw  \\
\lstick{Target \;\;}             \qw & \qw \qwbundle{n} &  \qw     & \qw                  & \qw        & \qw 
\end{quantikz} \end{adjustbox}
}
\sidesubfloat[] {
\begin{adjustbox}{scale=0.8}
\begin{quantikz}
\lstick{\text{Phase} \;\;\;\; }     \qw & \qw              & \qw       & \qw                  & \octrl{1}                      & \qw                      &  \qw \\
\lstick{\text{Selection} \;\; }        \qw & \qw \qwbundle{m} & \qw       & \gate{\text{Prep}}   & \gate[wires=2]{\text{Select}}  & \gate{\text{Prep}^{-1}}  &  \qw \\
\lstick{Target \;\;\;\; }           \qw & \qw \qwbundle{n} & \qw       & \qw                  & \qw                            & \qw                      &  \qw
\end{quantikz}\end{adjustbox}
}\hspace{20pt}
\end{center}
\caption{ {\bf Circuits for generalized quantum signal processing (GQSP).} (a) A circuit implementation of GQSP in the guise of a generalized quantum eigenvalue transform $\mathcal{O}_{\{\lambda,\phi,\theta\}}(H)$.  The signal processing rotations $R(\lambda,\phi,\theta)$ are operations in SU(2), which we decompose according to a $ZYZ$ convention.  (b) Structure of the walk operator $\mathcal{W}$ in terms of the block encoding $U_H$ and multicontrolled CZ gate. (c) Assembly of the block encoding $U_H$ using Select and Prepare oracles.  We use these circuit components to implement $H \mapsto \exp[-\imath H t]$.} \label{fig:qet_circuit_base}
\end{figure*}

\noindent giving access to each term $\Lambda_i$ in the LCU.  A  binary encoding $\ket{i}$ is used to flag $\Lambda_i$, meaning that $\ket{G}$ must be an $m = \lceil \log_2 M \rceil$ qubit state.  Similarly, we define a \emph{Prepare oracle},
\begin{equation}
U_\text{prep} = \frac{1}{\sqrt{\vert\vert c \vert\vert_1}} \sum_i \sqrt{c_i}\, \ket{i}\bra{G},
\end{equation}

\noindent which translates our selection state $\ket{G}$ into the weighted superposition required by the Select oracle.  Here $\vert\vert c \vert\vert_1 = \sum_i \vert c_i \vert$ is the 1--norm of the coefficient set $\{c_i\}$. This  serves as the normalization factor $\alpha = ||c||_1$ for our encoding. The block--encoded Hamiltonian is obtained by assembling these pieces:
\begin{equation}\label{eq:lcu_select_prepare}
U_H = (U_\text{prep}^\dagger \otimes I_n) \, U_\text{sel} \, (U_\text{prep} \otimes I_n).
\end{equation}


\subsection{Quantum Eigenvalue Transform}

\noindent  We generate our dynamics using a variant of QSP called the quantum eigenvalue transform (QET).  The QET sequence $\mathcal{O}_{\vec{\phi}}$ applies a polynomial function  across the spectrum of a normal operator,
\begin{equation}
\mathcal{O}_{\vec{\phi}}: H \mapsto f(H) = \sum_\lambda f(\lambda) \ket{\lambda}\bra{\lambda},
\end{equation}
\noindent which is defined by using a series of classically optimized phase angles $\{\phi_i\}$ to specify $f$  \cite{Low2017,Low2019,Gilyen2019,Martyn2021}.  For the purposes of quantum simulation, we would like to implement $f(H) = \exp[-\imath H t]$. A practical QET implementation requires the block encoding $U_H$, a qubitized reflection $Z_\Pi = 2 \ket{G}\bra{G} \otimes I_n - I_m \otimes I_n$ on the selection subspace, and a series of qubitized rotations $Z_{\Pi,\phi} = \exp[{\imath \phi Z_\Pi}]$. We assemble the transform as a product,
\begin{equation}\label{eq:qet_general}
\begin{split}
\mathcal{O}_{\vec{\phi}} &= e^{\imath\phi_0 \, Z_\Pi} \prod_{k=1}^d \left[U_H Z_\Pi e^{\imath\phi_k \, Z_\Pi}\right]. \\
\end{split}
\end{equation}
\noindent Note that the isolated reflection is often incorporated into  $U_H$ or handled through the phase convention.

\par While the QET has favorable asymptotic scaling, the use of $U(1)$ rotations can limit expressivity.  The leading obstacle is that we can only encode real or imaginary functions of definite parity. This means an evolution operator $x \mapsto \exp(-\imath x t) = \cos (xt) - \imath \sin (xt)$ must combine separate circuits for the sine and cosine as an LCU (e.g., following the symmetric  convention of \cite{Dong2021}).  Other restrictions further limit the transform to $H \mapsto \exp{(-\imath H t}/2)$. If the QET sequence gives in error of $\epsilon$ in the target function, the probability of a successful shot will be $p = \vert\vert \exp(-\imath H t)/2 \pm \epsilon \vert\vert^2$.  This is close to 1/4 for small $\epsilon$ and thus algorithm repetition is inevitable.  Robust oblivious amplitude amplification (ROAA) can increase the success probability to near unity \cite{Berry2015, Martyn2023} though this triples the algorithmic depth.

\par Generalized quantum signal processing (GQSP) \cite{Motlagh2024} is an alternative that follows from early QSP proposals \cite{Low2017,Low2019,Rines2019}. Here, the U(1) signal processing rotations are swapped for SU(2) operations,

\begin{equation}
R(\lambda,\phi,\theta) = \begin{pmatrix} e^{\imath (\lambda + \phi)} \cos(\theta) & e^{\imath \phi} \sin(\theta) \\ e^{\imath \lambda} \sin(\theta) & -\cos(\theta)\end{pmatrix}.
\end{equation}

\noindent This expanded set of of phases allows us to transform a block encoded operator $H$ as,

\begin{equation}
\left[ \prod_{k=1}^d R(0,\phi_k,\theta_k) A \right] R(\lambda_0,\phi_0,\theta_0) = \begin{pmatrix} P(\mathcal{W}) & \cdot \\ Q(\mathcal{W}) & \cdot \end{pmatrix},
\end{equation}

\noindent where $A = \ket{0}\bra{0} \otimes \mathcal{W} + \ket{1} \bra{1} \otimes I$ is defined in terms of a walk operator $\mathcal{W} = Z_\Pi U_H$.  The result is a pair of complex polynomials $\vert P(\mathcal{W})\vert^2 + \vert Q(\mathcal{W})\vert^2 = 1$ with fewer restrictions than conventional QSP. The GQSP formalism also specifies a simple and efficient algorithm to calculate  the required phase rotations and the form of one polynomial in terms of the other.  These benefits compound for Hamiltonian simulation and the  eigenvalue transform.  Here, the spectral representation of $\mathcal{W}= \bigoplus_\lambda W_\lambda$ has a convenient  form,
\begin{equation}
\mathcal{W} = \bigoplus_\lambda \begin{pmatrix} \lambda & \sqrt{1 - \lambda^2} \\ -\sqrt{1 - \lambda^2} & \lambda \end{pmatrix} \otimes \ket{\lambda}\bra{\lambda},
\end{equation}

\noindent with eigenvalues of $\exp(\pm \imath \cos^{-1} \lambda)$ in each qubitized eigenspace.  While this holds for any QET  variant, the generalized scheme also gives an efficient means to implement $P(\mathcal{W} ) = \exp(-\imath H t)$ as,
\begin{equation}
P(\mathcal{W}_\lambda)  = \begin{pmatrix} e^{-\imath t \cos(\cos^{-1} (\lambda))} & 0 \\ 0 & e^{-\imath t \cos(-\cos^{-1} (\lambda))}\end{pmatrix}.
\end{equation}

\noindent This is not possible with conventional  schemes  since $P(x)$ is a complex polynomial of indefinite parity.  An important consequence is that separate QET transformations are no longer required for sine and cosine components.  This eliminates the associated LCU and the need for amplification to boost the success probability. We adopt this generalization of QET for our time evolution circuits \footnote{Our practical use of the GQSP formalism actually corresponds to a quantum eigenvalue transform.  However, we retain the name GQSP to maintain consistency with literature.  However, it is currently unclear if GQSP can be extended to the more general context of the quantum singular value transform (QSVT).}.


\subsection{Phase Angle Generation}

\par The GQSP algorithm requires a set of phase angles that approximate $\mathcal{U}(t) = \exp(-\imath H t)$ to an error of $\epsilon$.  These are formally defined using a degree $d$ Chebyshev polynomial from the Jacobi--Anger expansion \cite{Low2017},
\begin{equation}
e^{i t \cos(x)} \approx \sum_{n=-d}^d \imath^n J_n (t) e^{\imath n x},
\end{equation}

\noindent where $J_n(t)$ is the $n$--th Bessel function of the first kind. Both the complimentary polynomial $Q(x)$ and the SU(2) phase angles $(\lambda_k, \phi_k, \theta_k)$ can be refined   using the algorithm of Ref.~\cite{Motlagh2024}.  This gives a series of $2d+1$ phase tuples and overhead comparable to a standard QET sequence (e.g., via \cite{Dong2021}). In particular, the full circuit will require $2d+1$ SU(2) rotations, $2d$ qubitized reflections, and $2d$ applications of our block encoding. Moreover, a  Jacobi--Anger expansion to degree $d = e|\tau|/2 + \log_{10}(1/\epsilon)$ gives a truncation error bounded by $\epsilon$.  The cited classical algorithm can refine up to $10^6$ phases in roughly one minute using readily accessible classical hardware and stated error thresholds. This assures us that  classical computations will not be a bottleneck since our small--molecule and protein estimates require less than $1.0 \times 10^7$ and $6.0 \times 10^7$ phases, respectively.


\subsection{Circuit Implementation}

\par  We use a straightforward implementation of GQSP that interleaves SU(2) rotations with controlled applications of the walk operator $\mathcal{W} = Z_\Pi U_H$.  Moreover, we alternate  $\mathcal{W}$ and $\mathcal{W}^\dagger$ between repetitions as required when encoding an expansion with terms of negative degree \cite{Motlagh2024}.  This arrangement is depicted in Fig.~\ref{fig:qet_circuit_base}a. The walk operator itself is constructed from the block encoding $U_H$ and a multicontrolled CZ gate that implements the qubitized reflection (Fig.~\ref{fig:qet_circuit_base}b).  Viewed differently, this leads to a multicontrolled SU(2) rotation between repetitions of $U_H$ or $U_H^\dagger$ similar to the qubitized rotation from conventional QET.  

\par The block encoding $U_H$ implements $H$ as an LCU of Pauli strings   using Select and Prepare oracles (Fig.~\ref{fig:qet_circuit_base}c). We adopt a Prepare that combines QROM lookup with an alias sampling strategy and which tolerates dirty ancilla outside of the selection register \cite{Babbush2018}.  Alias sampling approximates LCU coefficients to a fixed bit precision  using a space--time tradeoff that exchanges $T$--complexity for logical qubit overhead. This approach is particularly reasonable when inputs have a known experimental uncertainty.  Our Select is more conventional, applying operators based on binary iteration over the selection register.  Circuits are compiled to the universal Clifford+$T$ gate set before quantifying resources, as required for quantum error correction schemes like the surface code.  All explicit circuit manipulations  are accomplished using the pyLIQTR software suite \cite{pyLIQTRMS2024} and its Qualtran extensions \cite{Harrigan2024, QualtranSW2024}.  Our NMR workflow has also been made available as part of pyLIQTR.


\subsection{Observable Estimation and Sampling}

\par The proposed quantum simulations would be initiated using spin configurations that are polarized along the $z$-axis.  An  immediate approach is to runs simulations for ensemble of computational basis states that reflect a net positive magnetization (e.g., selected from eigenstates of $S^z_\text{tot}$).  The outputs can then be combined classically,
\begin{equation}
\mathcal{S}(t) = \sum_k \bra{\psi_{0,k}} \mathcal{U}^{-1}(t) S^z_\text{tot} \mathcal{U}(t)S^z_\text{tot}\ket{\psi_{0,k}}
\end{equation}
\noindent where $\mathcal{U}(t) =  \exp[-\imath H t]$ is the unitary time evolution operator to be approximated by QET.  Although the number of possible initial states grows exponentially in $N$, the variance in this estimator is much smaller. Rudimentary arguments show that $\mathcal{S}(t)$ can be well--reproduced by sampling at most $N^2$ of these computational basis states.  Achieving a measurement precision of $\epsilon_\text{meas}$ requires $O(1/\epsilon_\text{meas}^2)$ samples, so we presume a worst--case  of $O(N^2 / \epsilon_\text{meas}^2)$ shots. 


\begin{figure}
\begin{center}
\begin{adjustbox}{scale=0.75}
\begin{quantikz}
\lstick{\text{Readout} \;\;\;\;$\ket{0}$}   \qw & \qw               & \gate{H}                 & \octrl{}\vqw{4}                  & \qw                                   & \octrl{}\vqw{4}                & \qw                        & \qw \\
\lstick{\text{Phase} \;\;\;\; $\ket{0}$}    \qw & \qw               & \qw                      & \qw                              & \gate[3]{\mathcal{O}_H}               & \qw                           & \qw                        & \qw \\
\lstick{\text{Selection} \;\; $\ket{0^m}$}  \qw & \qw \qwbundle{m}  & \qw                      & \qw                              & \qw                                   & \qw                            & \qw                        & \qw \\
\lstick{Target \;\;\;\; $\ket{0^n}$}        \qw & \qw \qwbundle{n}  & \gate{\text{PUp}}        & \gate[2]{S^z_\text{tot}}         & \qw                                   & \gate[2]{S^z_\text{tot}}       & \gate{\text{PUp}^{-1}}     & \qw \\
\lstick{\text{Sum} \;\;$\ket{0^s}$}         \qw & \qw               & \gate{H^{\otimes m}}     &                                  & \qw                                   &                                & \gate{H^{\otimes m}}       & \qw 
\end{quantikz} \end{adjustbox}
\end{center}
\caption{
 {\bf Correlation function estimator.} Circuit estimating the correlation function $\langle S^z_\text{tot}(t) S^z_\text{tot}(0)\rangle$ based on phase kickback from total spin operators $S^z_\text{tot}$ to the single--qubit Readout register.  Time evolution is facilitated by the QET sequence $\mathcal{O}_H$ while the PUp prepare oracle is used to prepare a superposition over spin--up states.  The Hadamard transform $H^{\otimes m}$ does the same over all computational basis states, and serves as a prepare oracle for the binary encoding used by the total spin operator oracles.  The $n$--qubit state of interest is contained in the Target register while the Hamiltonian is prepared using the $m = \lceil\log_2 M\rceil$ selection register for the LCU.  An $s = \lceil\log_2 n\rceil$ register evaluates a sum of $Z$ Pauli operations over each qubit in the target state.  The total spin operator is constructed explicitly in Fig.~\ref{fig:total_spin_operator}. }
 \label{fig:estimator_circuit}
\end{figure}

\par An alternative is to prepare the initial states in superposition on the quantum computer.  This can be done by  generating a uniform superposition and filtering with inequality tests as in Ref.~\cite{Babbush2018}.  The correlation function $\mathcal{S}(t) = \langle S^z_\text{tot}(t) S^z_\text{tot}(0)\rangle$ is then extracted using phase kickback from a pair of total spin operators followed by amplitude estimation (Fig.~\ref{fig:estimator_circuit}).  A simple amplitude estimation strategy might use the Hadamard test with a $O(1/\epsilon_\text{meas}^2)$ overhead in the number of shots that estimate $\mathcal{S}(t)$.  A robust error threshold would be on the order of $\epsilon_\text{meas} = 0.01$ for a total of $N_\text{shots} = 10^4$ shots, while a more permissive value of $\epsilon_\text{meas} = 0.05$ gives a markedly reduced $N_\text{shots} = 400$ shots.  Iterative phase or amplitude estimation procedures are an alternative, though they require repetition of the overall circuit \cite{Grinko2021}. A nuanced discussion can be found in Ref.~\cite{Agrawal2024}.

\par Another consideration involves the set of timepoints used to reconstruct $\mathcal{S}(t)$ and thus the target spectrum $\mathcal{S}(\omega)$.  A na\"{i}ve approach might uniformly sample $\langle S^z_\text{tot}(t) S^z_\text{tot}(0) \rangle$ with a timestep $\Delta t$  so that $N_\text{points} = t_\text{max} / \Delta t$ is much greater than $N$.  Based on reference experiments, at least $2^{12}$ points might be required for a high--resolution spectrum.  However, the  sparse nature of $\mathcal{S}(\omega)$ is a prime candidate for compressed sensing \cite{Donoho2006}, which can effectively reduce timepoint samples in NMR experiments \cite{Bostock2017, Bostock2018, Robson2019}. This improvement ranges from a factor of two for high--precision spectra \cite{Delaglio2017} to a factor of 40 for crude spectra from  NISQ algorithms \cite{Seetharam2023}.  Again, our goal is to reproduce the location of resonances and not to mimic high--accuracy data.  We will assume that $N_\text{points} = 400$ is suitable for most purposes.

\par  Putting this together, we must execute $N_\text{shots} \times N_\text{points}$ time evolution circuits to predict a spectrum.  This ranges between $4 \times 10^4$ and $4 \times 10^6$ shots depending on the target application.  However, these points correspond to a range of simulation times. Many of these will be far shorter than the maximal simulation time, with a commensurately lower resource overhead.


\begin{figure*}
\begin{center}
\includegraphics[width=1.0\columnwidth]{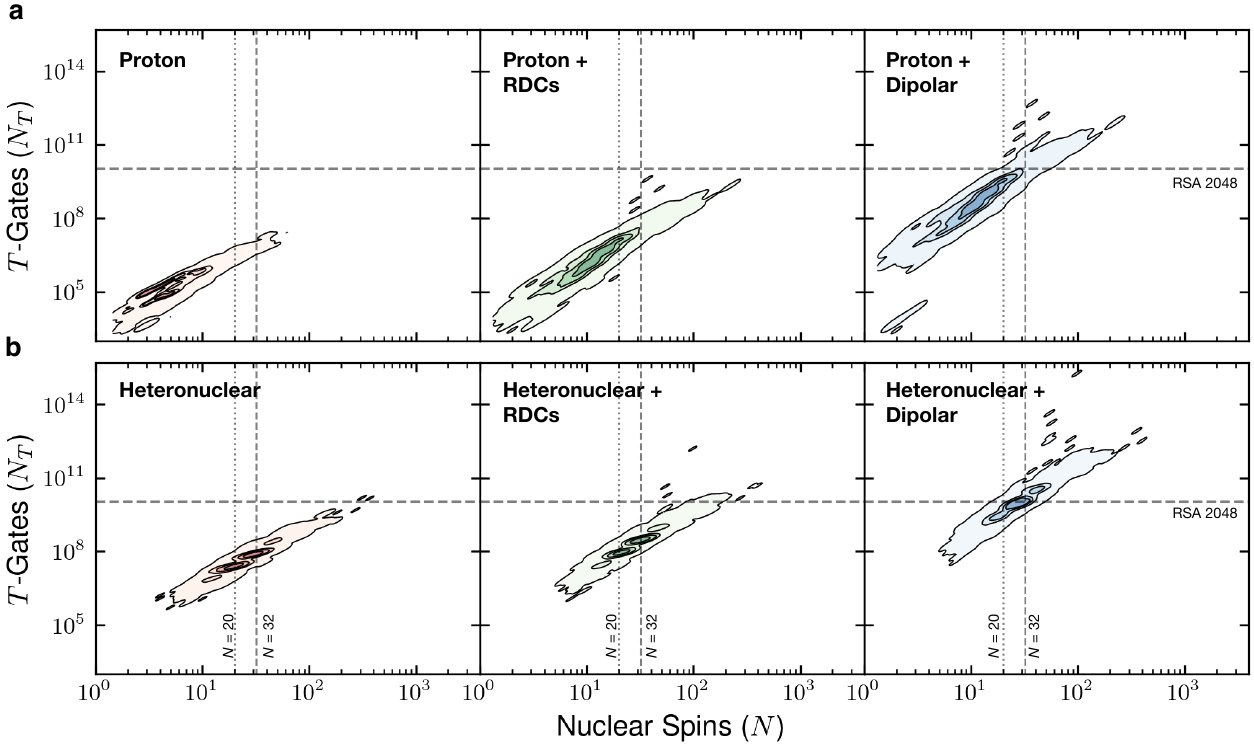}
\end{center}
\caption{ {\bf Limiting overhead vs. problem scales for small--molecule NMR simulations.} Single--shot $T$--gate count ($N_T$) for simulating the largest spin cluster of a molecule up to $t_\text{max} = 1\,\text{s}$ with nuclear spin networks of size $N$.  Datasets correspond to Hamiltonians with $J$--couplings between magnetically active protons ({\bf Proton}; $^1$H) and (b) a broader set of nuclei ({\bf Heteronuclear}; $^1$H, $^{13}$C, $^{15}$N).  Panels in each row correspond to these cases or the addition of residual ({\bf + RDCs}) or full ({\bf + Dipolar}) dipolar couplings. A maximal error of $\epsilon = 5 \times 10^{-3}$ is budgeted when approximating the evolution unitary with GQSP, assuming a spin--spin relaxation time of $T_2 = 1$ s.  Contours are kernel density estimates for $1.2 \times 10^4$ different molecules, using 15 isovalues thresholded to enclose all points.  The horizontal line is the estimated $1.08 \times 10^{10}$ $T$--gates required to factor a 2048--bit integer in 8 h using Shor's algorithm \cite{Gidney2021}. Vertical lines are thresholds for classical hardness based on systems with $N = 20$ and $N = 32$ spins.}
\label{fig:spins_vs_t_complexity}
\end{figure*}


\section{Resource Estimates}

\par Resource estimates often use a small set of inputs and assume that they  generalize to a larger problem class.  This strategy is typically viable, though  it can fail when there are large variations in problem structure between instances. To avoid biasing our estimates, we have generated explicit time--evolution  circuits for a diverse set of roughly $1.2 \times 10^4$ small organic molecules and 211 biological macromolecules \cite{Carr2005}.

\par We quantify temporal overhead using the logical $T$--gate count ($N_T$) for the longest time--evolution circuit $t = t_\text{max}$ of the largest spin cluster used to calculate $\langle S^z_\text{tot}(t) S^z_\text{tot}(0)\rangle$. We also use the number of nuclear spins $N$ as a first--order metric for  problem complexity. Spatial overhead is quantified through the number of logical qubits $N_L$ in the aforementioned circuit.  This includes the $N$ qubits that represent system as well as ancilla for the block encoding, dynamics, and observable extraction. 

\par Qubit count is generically a poor proxy for overall computational complexity.  For instance, an identity circuit can be applied to an arbitrary number of qubits with high spatial overhead but low temporal complexity.  However, our nuclear spin networks have universal features that increase complexity in concert with $N$.   These estimates are contextualized by specifying hardness scales at $N = 20$ and $N = 32$ spins, which are justified further in Appendix~\ref{appendix:utility_thresholds}. In particular, na\"{i}ve classical algorithms will find explicit application of the time--evolution operator cumbersome at $N = 20$   and could be aided by a fast quantum computer \footnote{At $N = 16$, a matrix representation of the time--evolution operator would be 64 Gb (assuming 64--bit complex entries).  This would be time--consuming to integrate on the timescales required for simulating spectra.  At $N = 20$ the operators would require 16384 Gb matrices, which is prohibitive for any classical calculation.}. Similarly, entanglement growth can make accurate matrix product state (MPS) methods with long--range interactions costly at $N = 32$.  It is important to note that our goal is not quantum advantage but instead practical utility. That is, a fast quantum computer that gives meaningful speedup will be of value even if the problem is formally tractable.


\subsection{Small--Molecule Spectra}

\par Magnetically silent nuclei can disrupt scalar coupling networks to give disconnected spin clusters. The largest cluster defines a limiting simulation  complexity, assuming that features of the ensemble are roughly scale invariant. Motivated by this, we extracted the largest clusters under each coupling Hamiltonian for all small molecule instances. The single--shot logical overhead for these is presented in Fig.~\ref{fig:spins_vs_t_complexity}.   Remarkably, the overhead for simulating these up to $t = t_\text{max}$  requires less than $10^{10}$  $T$--gates in the $J$--coupled {\bf Proton} or {\bf Heteronuclear} cases with scalar couplings. This is also true when residual dipolar coupling ({\bf +RDCs}) terms are added to give larger and more connected spin clusters. The cited bound is notable since it demarcates the previously estimated logical overhead for factoring 2048--bit integers in 8 h with Shor's algorithm and a rotated surface code \cite{Gidney2021}. Quantum algorithms are provably advantageous and potentially consequential in this context, making the factoring result a ``standard candle'' for utility and computational scale.  Hamiltonians with full dipolar couplings can surpass this overhead ({\bf +Dipolar}), but only by a maximal factor of $10^2$.  While these have almost the same number of terms as their {\bf +RDC} counterparts, the larger coupling strengths markedly increase the normalization $\alpha$ and thus the length of the QET sequence.   All small--molecule instances  require less than 300 logical qubits, which is below the 6190 logical qubits needed for factoring.

\par
\begin{figure}
\begin{center}
\includegraphics[width=1.0\columnwidth]{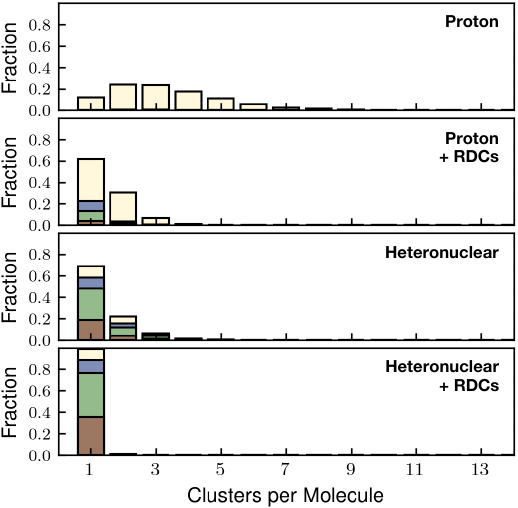}
\end{center}
\caption{ {\bf Nuclear spin cluster statistics.} Yellow bars denote the fraction of our small--molecule dataset containing a given number of disconnected nuclear spin clusters. The coupling regime is labeled in bold. Blue, green, and brown bars are fractions of the dataset with $N \geq 16$, $N \geq 20$ and $N \geq 32$ spins in the largest cluster, respectively.   }
\label{fig:cluster_distributions}
\end{figure}

 While these estimates are suggestive, it is important to delineate problems that will truly benefit from quantum computation.  As a general rule, large, highly--connected clusters can be challenging for classical algorithms, particularly when they contain numerous couplings of comparable magnitude.  However, there is a regime where these remain tractable using Liouville--von Neumann simulators \cite{Hogben2011}, tensor networks, or other methods for highly correlated spin Hamiltonians \cite{Menicucci2002, Appelt2010}.  We capture this feature for our nuclear spin clusters in Fig.~\ref{fig:cluster_distributions}. 

\par Our simplest spin networks are defined by scalar couplings between protons ({\bf Proton})  that are separated by up to four bonds. This leads to small, isolated spin clusters that are largely amenable to classical computation.   Dipolar couplings modify this by linking spatially proximate protons to give higher connectivity.  This is apparent in the {\bf + RDCs} panels of Fig.~\ref{fig:cluster_distributions}, which are characterized by fewer and larger clusters in a greater proportion beyond  classical hardness thresholds.  Heteronuclear $J$--couplings have a similar effect since they deliver  additional  coherence pathways between bonded atoms.  Dipolar couplings compound this connectivity to give a single, molecule--spanning spin cluster. In this limit, over 90\% of our instances should be challenging for   at least one classical algorithm.


\begin{figure}
{\footnotesize
\begin{center}
\includegraphics[width=1.0\columnwidth]{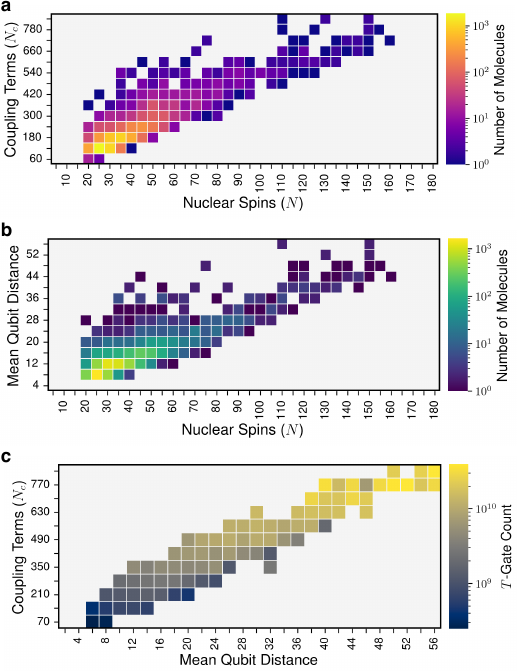}
\end{center}
}
\caption{{\bf Properties of nuclear spin Hamiltonians based on heteronuclear $J$--couplings.}  Distributions are presented for the number of molecules with given a nuclear spin count $N$ and (a) number coupling terms $N_c$ in the Hamiltonian or (b) mean distance between coupled qubits in a one--dimensional register.  The single--shot $T$--gate count for simulations up to $t = t_\text{max}$ is also presented in (c). Data correspond to molecules with $N \geq 20$ spins in the {\bf Heteronuclear} coupling regime, which includes $^1$H, $^{13}$C, and $^{15}$N nuclei.   Resource estimates are from Fig.~\ref{fig:spins_vs_t_complexity}.}\label{fig:molecular_ham_properties}
\end{figure}

\par Cluster size can be proportional to classical hardness, but this it is not guaranteed.   The exponential scaling of Hilbert space dimension with spin count will limit the  na\"{i}ve classical algorithms, though sparse simulators and tensor network methods can surpass full--rank simulations for many problems.  To further constrain quantum utility, we consider additional metrics including number $N_c$ of spin--spin coupling terms in a Hamiltonian and the distance $q_{ij}$ between the coupled qubits in a one--dimensional register. These capture  the sparsity of matrix Hamiltonians and the potential to generate long--range entanglement.  The latter can be especially prohibitive for 1d tensor networks like matrix product states (MPS), where the numerical overhead scales as $O(e^S)$ in terms of the bipartite entanglement entropy $S$ between constituent tensors.  While a judicious state ordering can mitigate this in some cases \cite{Rams2020, Wojtowicz2020, Wojtowicz2021, Elenewski2021}, it is unlikely for spin networks with dipolar couplings.  

\par In Fig.~\ref{fig:molecular_ham_properties}, we present a distribution of these quantities for molecules with heteronuclear and homonuclear scalar couplings.  These  spin networks constitute a single molecule spanning cluster.  The typical instance has many couplings per spin (Fig.~\ref{fig:molecular_ham_properties}a) and an ensemble--averaged coupling degree of ${\langle\langle N_c / N \rangle\rangle} = 9.0$.  This is accompanied by a mean distance of $\langle\langle d_{ij}\rangle\rangle = 12.6$ between qubits representing coupled spins in our one--dimensional register (Fig.~\ref{fig:molecular_ham_properties}b), and an mean maximum distance of $\langle\langle \max_{ij} d_{ij}\rangle\rangle = 30.5$ qubits.  While MPS evolution algorithms can treat long--range couplings up to $N \sim 50$ spins with some efficiency, these states are often ordered so the coupling strength decays algebraically with distance.  That is not the case for our nuclear spin Hamiltonians, where the numerous long--range couplings  are often comparable in strength to their local counterparts (or  up to $10^3$--fold larger for full dipolar couplings in the solid--state).  These parameters lie near the mode of the distribution, as 43\% of molecules have spin networks that exceed the mean node degree and 39\% exceed the mean distance. Based on this, it is likely that a substantial fraction of molecules would see a benefit from quantum computation when some dipolar couplings are present. It is significant that these parameters generally increase with the spin count $N$ and are positively correlated with a greater $T$--gate count, Fig.~\ref{fig:molecular_ham_properties}c.  Thus, classical and quantum hardness are expected to scale in concert for this problem.

\par So far, we have only considered the overhead for a single quantum dynamics simulation. However, a major difference between number theoretic algorithms and the estimation of a correlation function is the need to consider a multitude of timepoints. We must also simulate all classically hard spin clusters in a given molecule.  The latter will not increase the maximal number of required qubits since the estimates of Fig.~\ref{fig:spins_vs_t_complexity} already correspond to the largest spin cluster.  In fact,  nearly all instances from the {\bf Heteronuclear + RDCs} case comprise a single cluster that spans across the entire molecule \footnote{While these clusters span the whole molecule, they are not saturated in the number of Hamiltonian terms (and thus $T$--gate count). This is due to our finite range for dipolar interactions, which is smaller than the maximal extent of the molecule. Stated differently, these models have a fully connected spin network, but do not saturate the edge count.}. The single--shot $T$--gate count for the leading cluster is also bounded since these estimates correspond to the longest simulation time $t_\text{max}$ (though this does not have to bound the maximal $T$--gate count for individual clusters).


\begin{figure*}
\begin{center}
\includegraphics[width=1.0\columnwidth]{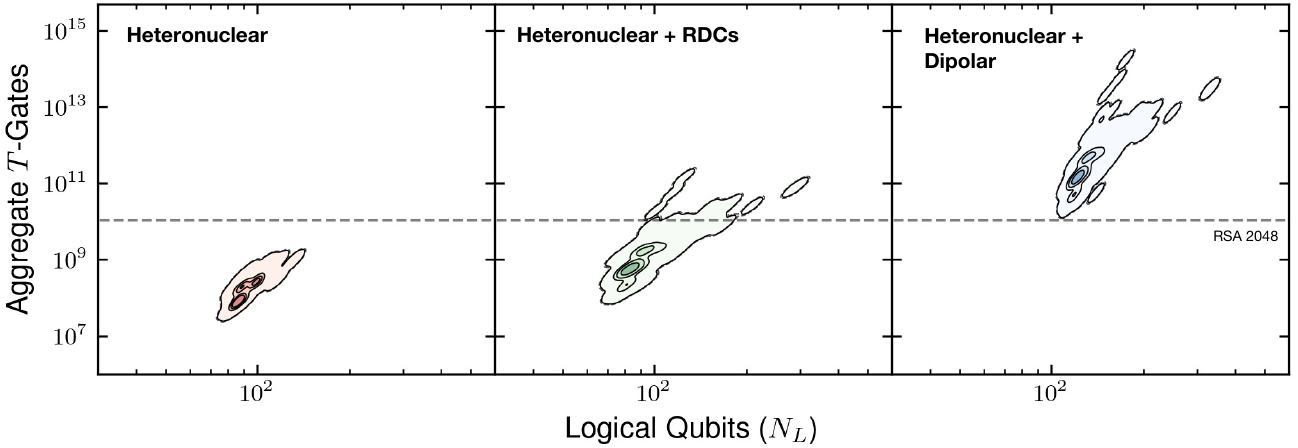}
\end{center}
\caption{ {\bf Aggregate single--shot overhead for small--molecule NMR simulations.} Estimates of the aggregate $T$--gate count required to perform a single--shot dynamics simulation for all quantum--relevant clusters ($N \geq 20$ spins) over $N_\text{step} = 400$ logarithmically spaced timepoints up to $t = t_\text{max}$. Parameters and precision targets match  Fig.~\ref{fig:spins_vs_t_complexity}.  Datasets contain 156 molecules ({\bf Heteronuclear}), 1806 molecules  ({\bf + RDCs}) and 1806 molecules ({\bf + Dipolar}), respectively. These  are smaller than the full dataset since some molecules lack clusters beyond the $N \geq 20$ threshold. }
\label{fig:aggregate_spins_vs_t_complexity}
\end{figure*}

\par A useful metric is the aggregate overhead for a single shot of all calculations (clusters and timepoints) used in estimating a nuclear spin correlation function.  That is, we estimate $T$--gate counts for running $N_\text{step}$  simulations of duration $t_k \in [0,t_\text{max}] $ and  aggregate these over a set of $M$ relevant clusters: $N_{T,\text{agg.}} = \sum_{k=1}^{N_\text{step}} \sum_{m=1}^{M} N_{T}(m, t_k)$.  Since we can use a compressed sensing protocol, it is likely that the time samples will be weighted toward early points.
 We approximate this by using a  logarithmic discretization of the time domain so that $ t_k \in [1/(2 f_\text{max}), t_\text{max}]$.  The frequency $f_\text{max}$ is an upper bound on the fastest coupling in the spin network and the factor of two is added to satisfy Nyquist requirements.  Since small clusters can be treated classically, we only include those with $N \geq 20$ in our estimates.  This is beyond the scale of exact simulation, though we presume that some approximate method (e.g., tensor networks or otherwise) will have classical utility near or below this scale. Again, a sufficiently fast quantum computer might have utility even when a cluster is classically tractable. These assumptions lead to the estimates of Fig.~\ref{fig:aggregate_spins_vs_t_complexity}.  On the single--shot level, we see that the aggregate $T$--gate count exceeds our longest simulation of the largest cluster in each molecule by roughly a factor of $10^2$.  The number of required shots $N_\text{shots}$ will depend on the target measurement precision $\epsilon_\text{meas}$ and protocol as discussed earlier.


\subsection{Biological Macromolecules}
 
\par
\begin{figure}
{\footnotesize
\begin{center}
\includegraphics[width=1.0\columnwidth]{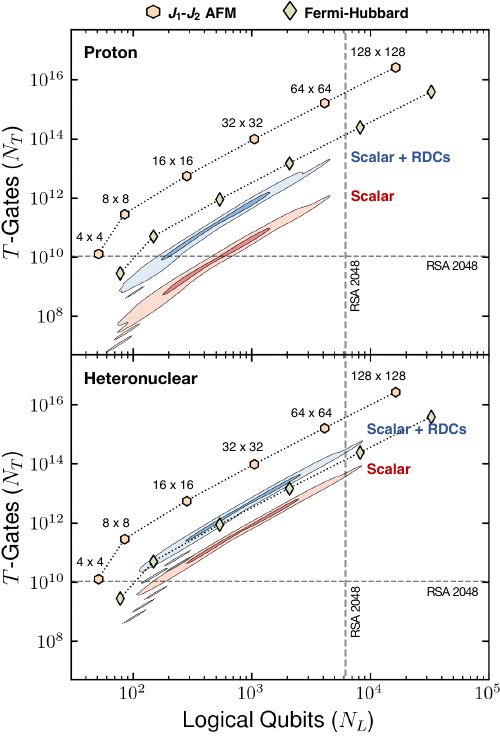}
\end{center}
}
\caption{{\bf Overhead estimates for simulating biomolecular (protein) spin networks.} Single--shot $T$--gate ($N_T$) and logical qubit ($N_L$) counts for simulating 211 different protein spin networks up to $t_\text{max} = 1\,\text{s}$.  Hamiltonians are defined for the largest spin cluster in each regime, which is generally a single cluster than spans the entire protein. Parameters match those of Fig.~\ref{fig:spins_vs_t_complexity} and estimates are presented as a kernel density at three contour levels.  Reference estimates are also shown for the single--orbital Fermi--Hubbard model on a square lattice  (green diamonds) and a $J_1-J_2$ Heisenberg antiferromagnet on a triangular lattice (yellow hexagons).  Lattice Hamiltonians and parameters are described in Appendix~\ref{appendix:condensed_matter_models}.}\label{fig:biomolecular_overhead}
\end{figure}

 Our estimates suggest that quantum computation might have sub--Shor utility for simulating the nuclear spin networks of certain small organic molecules. Macromolecular spin networks are even more taxing for classical computation and they have greater spectroscopic utility \cite{Sprangers2007,Weingarth2013,Rosenzweig2016,Puthenveetil2019,Bayer2020,Reif2021,Hu2021,Hu2021, Marusic2023, Fontana2023}. To assess this case, we have examined a series of proteins within the {\bf Proton} and {\bf Heteronuclear} coupling regimes with optional residual dipolar couplings ({\bf + RDC}). This macromolecular dataset includes drug--discovery relevant peptides as well as larger protein structures from a previously designed benchmark \cite{Klukowski2024}.   Here we find a tight correlation between problem qubit $N$ and $T$--gate counts (Fig.~\ref{fig:biomolecular_overhead}). This reflects the compact spatial conformations of folded proteins and the fact that these biopolymers assembled from a fixed pool of amino acid  blocks.  Interestingly, all molecules in this dataset still lie below the logical qubit count for 2048--bit integer factoring.  The largest proteins that we consider are ion channels -- the P7 channel and the human voltage--dependent anion channel (VDAC) -- and a maltodextrin binding protein. All three are challenging for NMR experiments and have no feasible path toward classical computation in the ZULF regime.

\par We contextualize this overhead by comparing to key quantum dynamics problems, including a $J_1$--$J_2$ Heisenberg antiferromagnet on a triangular lattice and the single--orbital Fermi--Hubbard model on a square lattice (see Appendix~\ref{appendix:condensed_matter_models} for details).  These are effective models for spin--liquids and strongly--correlated electronic materials and large instances of both remain intractable to classical computation for large instances. More specifically, $16 \times 16$ site lattices lie near the envelope of current classical computations, though unambiguously valuable information should become apparent at  $128 \times 128$ lattices and beyond.  Our estimates suggest that simulating protein NMR spectra has temporal complexity comparable to reproducing correlation functions for these condensed matter systems.  Moreover, this can be done with fewer logical qubits.  
\par Our analysis has assumed that molecules are at most weakly aligned.  Realistic  solid--state NMR predictions would require longer distance cutoffs for dipolar interactions. Moreover, individual dipolar couplings would be roughly  1000--fold larger than the RDC counterparts, substantially increasing the normalization factor $\alpha$ and thus the effective simulation cost.  This would give a markedly elevated $T$--gate count, though the spatial resource utilization would be comparable.


\subsection{ Utility and Physical Resource Overhead}

\par
\begin{figure}
\begin{center}
\includegraphics[width=1.0\columnwidth]{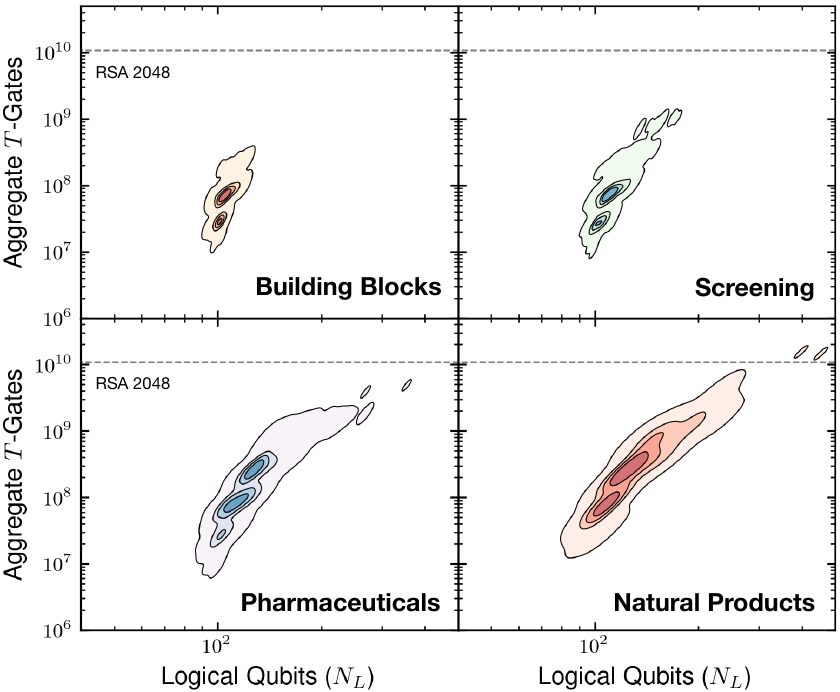}
\end{center}
\caption{ {\bf Categorical distribution of single--shot overhead for small--molecule NMR simulations.} Estimates of the aggregated $T$--gate count needed to perform a single--shot dynamics simulation for all quantum--relevant spin clusters ($N \geq 20$ spins) at $N_\text{step} = 400$ logarithmically spaced timepoints up to $t = t_\text{max}$. Data, parameters and precision targets match those of Fig.~\ref{fig:spins_vs_t_complexity}.   }
\label{fig:aggregate_spins_vs_t_complexity_categorical}
\end{figure}

  ZULF NMR is an emerging technique with potential impacts throughout chemistry and materials science.  Our input molecules are drawn from real--world datasets, allowing us to speculate on the value of quantum computation in supporting this spectroscopy.  In Fig.~\ref{fig:aggregate_spins_vs_t_complexity_categorical} we present a categorical classification of our small--molecule estimates with scalar heteronuclear couplings.  From a practical standpoint, only larger pharmaceuticals or natural products are very likely to remain inaccessible with foreseeable improvements in classical algorithms.  This suggests that the strongest candidates for quantum utility would be spectroscopically--challenging small--molecules or macromolecules. The former might uniquely benefit from ZULF spectroscopy since high--field experiments can struggle to resolve an unambiguous structure.  Rough estimates for economic impact from these calculations are provided in Appendix~\ref{app_economic_utility}.

\par Since the proposed calculations would facilitate experiments, their utility is inherently tied to their runtime.  We have estimated this for spectroscopically--challenging reference molecules and therapeutic--scale peptides in Table~\ref{table:physical_overhead}. This analysis assumes a planar layout of spin or superconducting qubits with nearest--neighbor couplings and a quantum error correction layer based on lattice surgery and the rotated surface code (detailed in Appendix~\ref{appendix:physical_estimate_methodology}).  While simulations with the minimal number of problem qubits would be prohibitively slow, a quantum computer that could factor 2048--bit integers (6190 logical qubits) or simulate a $128 \times 128$ site instance of the Fermi--Hubbard model (32805 logical qubits) allow sufficient parallelization to be useful.  However, a quantum computer could also be  advantageous for even smaller molecules if it is sufficiently fast.  An interesting takeaway from this analysis is the interplay between the Hamiltonian regime, molecular scale, and physical overhead.  Notably, small molecules with heteronuclear couplings can surpass the quantum resource requirements for simulating drug--like peptides with a simple spin Hamiltonian. These systems are likely to have higher utility and value should increase with the size of a protein.  We do not press this analysis further since larger molecules will likely to require a greater number of timepoints due to spectral crowding. 

\par We have provided estimates based on the surface code due to its maturity and ubiquity.  Other error correction schemes such as quantum low--density parity--check codes (qLDPCs) and yoked surface codes can offer more robust encoding rates and potentially lower resource requirements \cite{Breuckmann2021, Bravyi2024, Gidney2025b}.  This has been established in revised estimates for factoring problems \cite{Gidney2025}, though the formalism and architectures for these alternatives are not universally well developed. 


\begin{figure*}
\begin{center}
\setlength\tabcolsep{0pt}
\begin{tabular*}{\linewidth}{ @{\extracolsep{\fill}} lcccccrrr  }
\hline\hline
 Molecule     &  $N$ & $N_L$ & $(d_2,d_1)$ & $\mathcal{N}_\text{fact}$ & $N_\text{phys}$        & Minimal Qubits & Gidney--Eker\aa & FH 128  \\ 
\hline 
{\bf Proton + RDCs}    &      &       &                         &         &       &       \\
\hline
IPC & 18 & 74 & (13, 7) & 4 & $ 1.8\times 10^{5}$ & 4294.0 h & 39.4 h & 16.4 h  \\
Parthenolide & 20 & 76 & (13, 7) & 4 & $ 1.8\times 10^{5}$ & 4743.3 h & 43.9 h & 18.2 h  \\
Strychnine & 22 & 78 & (13, 7) & 4 & $ 1.9\times 10^{5}$ & 3300.7 h & 30.8 h & 12.8 h  \\
\hline
{\bf Heteronuclear}     &       &       &                         &         &       &       \\
\hline
IPC & 27 & 106 & (15, 9) & 4 & $ 2.9\times 10^{5}$ & 32286.0 h & 467.9 h & 193.3 h  \\
Parthenolide & 35 & 114 & (15, 9) & 4 & $ 2.9\times 10^{5}$ & 44630.3 h & 666.1 h & 273.8 h  \\
Strychnine & 45 & 124 & (15, 9) & 4 & $ 3.0\times 10^{5}$ & 64655.9 h & 994.7 h & 409.2 h  \\
\hline
{\bf Proteins (Proton)}     &       &       &                        &         &       &       \\
\hline
$\alpha$--Conotoxin & 99 & 168 & (13, 7) & 4 & $ 2.5\times 10^{5}$ & 7205.9 h & 91.2 h & 37.7 h  \\
Gramicidin & 124 & 193 & (13, 7) & 4 & $ 2.7\times 10^{5}$ & 7620.0 h & 103.0 h & 42.8 h  \\
Mastoparan & 126 & 195 & (15, 9) & 4 & $ 3.7\times 10^{5}$ & 18719.8 h & 353.2 h & 145.1 h  \\
\hline\hline
\end{tabular*}
\end{center}
\caption{ {\bf Physical overhead for spectroscopically--challenging molecules.}   Runtimes reflect the concurrent execution of algorithm repetitions on systems  (i) with the minimal physical qubits for simulation, (ii) capable of factoring 2048--bit integers in 8 h  (Gidney--Eker\aa; 6190 logical / 20 M physical qubits) and (iii) capable of simulating the single--orbital Fermi--Hubbard model a $128 \times 128$ lattice (FH 128; 32805 logical / 48.1 M physical qubits).  Estimates are based on our explicit logical counts and a fast lattice surgery layout using the rotated surface code. We assume an error budget of $\epsilon_\text{target} = 0.1\%$ for the QEC layer, measurement depth $\mathcal{D}_\text{meas} = N_T$, cycle time of $t_\text{cycle} = 1$ $\mu$s, reaction time $t_\text{react} =  10$ $\mu$s, and an error rate of $0.01\%$ for physical two--qubit gates. $T$--factory distances $(d_2, d_1)$ and counts $\mathcal{N}_\text{fact}$ are optimized to meet this budget, giving physical qubit counts $N_\text{phys}$ and runtimes following the methodology of Appendix~\ref{appendix:physical_estimate_methodology}. The code distance $d = d_2$ of data tiles is matched to our level--2 factories. }
\label{table:physical_overhead}
\end{figure*}


\section{Conclusions}

\par We have analyzed how fault--tolerant quantum computation might augment  certain low--field NMR spectroscopies.  A unique aspect of this work is that it addresses a large number of inputs and provides explicit, circuit--based estimates using state--of--the-art qubitized algorithms (following \cite{Rines2019}). Based on our estimates, several classes of molecules remain at or beyond the limits of classical computation yet feasible for quantum computation. For example, our resource estimates suggest that small-molecule NMR spectra can be simulated with fewer than $10^{10}$ $T$--gates, a computational burden comparable to factoring $2048$-bit integers using Shor’s algorithm (though recent estimates have pushed this to lower resource counts \cite{Gidney2025}). Macromolecular systems with up to 124 spins in their largest cluster require fewer than $10^5$ logical qubits and $T$--gate counts comparable to those required for condensed--matter models like $128 \times 128$ Fermi--Hubbard lattices.  This positions ZULF NMR spectral prediction as a high--impact application for early fault--tolerant quantum hardware, with significant implications for spectroscopy--driven disciplines such as chemistry, biology, and materials science.

A quantum computer can also have utility for classically tractable tasks if it is sufficiently inexpensive and  high throughput. This is particularly true for  NMR spectral predictions, which have not enjoyed large, dedicated high--performance computing allocations like other frontier--scale chemistry and physics problems.  The stated limitations are not just due to resource availability.  Since spectral predictions should compliment individual experiments, a task that requires weeks of compute time  could have be  diminished in utility for many small molecules (though still be relevant for challenging targets like natural products and peptides).  Conversely, the ability to simulate spectra for large numbers of molecules could have added value by virtue of computational throughput.

\par Our effort follows several discussions of NMR simulation in a NISQ context \cite{Seetharam2023}.  Beyond low--field instruments, it has been proposed that quantum computation can find utility predicting macromolecular and solid--state materials spectra for high--field solid--state NMR.  This is certainly possible based on our observations.   However, the large number of strong, long--range interactions will markedly increase the block encoding normalization $\alpha$ and linearly increase the runtime linearly by the same factor.  The cost would be even greater in a high--field regime, where the nonzero chemical shift terms $\omega_{k0}$ can be a factor of $10^6$ larger than other couplings.  Thus, the most plausible use case for solid--state spectra remains in the ZULF regime.  Another opportunity for quantum simulation involves  the design of  pulse sequences for spectrometers \cite{Rule2006}.  This is inherently a quantum control problem, and the ability to tackle large spin systems could lead to unanticipated experimental techniques.  However, both this and solid--state NMR require methods for the  simulation of explicitly time--dependent Hamiltonians \cite{Chen2021,An2022}.

\par As a final note, we have steered toward pessimistic estimates for resource overhead. This is particularly true when considering spectra with heteronuclear couplings.  However, many experiments are performed for homonuclear proton networks, with  select heteronuclei included through isotopic enrichment.  This means that practical Hamiltonians would contain fewer terms with a commensurate reduction in $T$--complexity.


\section{Acknowledgements}

This material is based upon work supported in part by the Defense Advanced Research Projects Agency (DARPA) under Air Force Contract No. FA8702-15-D-0001. In addition, this research was developed in part with funding from the DARPA  under Contract No. HR00112330014.  Any opinions, findings, conclusions or recommendations expressed in this material are those of the author(s) and do not necessarily reflect the views or policies of the Defense Advanced Research Projects Agency, the Department of Defense or the U.S. Government


\section{Appendices}


\subsection{Dataset Curation and Input Preparation}	

\par Small--molecule structures were prepared using classical, molecular mechanics based methods.  Initial all--atom geometries were generated from a standard library of bond angles, as defined by the molecular connectivity and resulting orbital hybridization.  These preliminary geometries were relaxed using conjugate gradient minimization according to the Merck molecular mechanics force field (MMFF94) \cite{Halgren1996}. No electrostatic or van der Waals cutoffs were applied, and optimizations were conducted to a relative energy gradient of $1 \times 10^{-8}$.  This strategy is sufficient for many (non--reactive) biomolecular simulations, where the high environmental temperature often permits a classical description of molecular geometry.  Numerical optimizations were handled using the Open Babel toolkit \cite{OBoyle2021}.  Small molecule inputs were drawn from Maybridge libraries (Building Block, RO3 Diversity, and Screening collections), pharmaceuticals from non--biologic drugs approved by the United States Food and Drug Administration (the FDA ``Orange Book''), and natural products were curated from PubChem databases \cite{Maybridge, Kim2025, OrangeBook}. 

\par Protein inputs are based on solution NMR structures, as drawn from the protein databank (PDB) \cite{RCSB} and established reference sets \cite{Klukowski2024}. No further optimizations were applied since these geometries are realistic targets for the proposed quantum algorithm workflow.


\subsection{Classical Computation and Quantum Utility Thresholds}\label{appendix:utility_thresholds}

\par Quantum utility is a moving target that is defined, in part, by the frontier of classical computation.  In this section, we briefly remark on prior work and leading classical algorithms that guide our hardness thresholds.

\par Classical simulations of ZULF spectra have generally  utilized full--rank density matrix evolutions and a judicious choice of basis to achieve a per--timestep complexity of $O(2^{2N})$ in problem size $N$ \cite{Wilzewski2017}. These employ near--optimal matrix multiplication algorithms to go beyond a $O(2^{3N})$ scaling  but are less efficient than sparse statevector simulators and tensor network methods.  To put these methods in perspective, a system of $N = 20$ spins requires 16384 Gb to store the full time--evolution operator or a density matrix (assuming 64--bit complex entries).  While formally tractable, this requires the resources of a small cluster and makes integration up to relevant timescales cumbersome.

\par Tensor networks --- such as matrix product states (MPS) ---  offer a systematic tradeoff between  accuracy and computational cost \cite{Paeckel2019, Cirac2021}.  MPS factorizations are driven by the singular value decomposition (SVD) and  reduce the overhead to store a statevector from $O(2^N)$ to $O(d N\chi^2)$ using iterative truncations that are optimal in the Frobenius norm (similarly for operators). Here $d = 2$ is the dimension of our computational basis and the bond dimension $\chi$ reflects the maximal number of  classical parameters that are retained between truncations.  This ansatz associates a tensor $A^{j_\alpha}_{s_{\alpha-1} s_\alpha}$ with each physical site $\alpha \in \{1 \ldots N\}$, where states are index by $j_\alpha \in \{0,1\}$ in our  $d=2$ computational basis.  The tensors are contracted by summing over indices $1 \leq s_\alpha \leq \chi$ to give a compressed statevector:
\begin{multline}
\ket{\Psi} = \sum_{j_1j_2\ldots j_N} C_{j_1j_2\ldots j_N} \ket{j_1}\otimes\ket{j_2} \dots\ket{j_N} \\
 \approx \sum_{j_1j_2\ldots j_N, \{s\}} A^{j_1}_{s_1} A^{j_2}_{s_1 s_2}\ldots A^{j_N}_{s_{N-1}}\ket{j_1}\otimes\ket{j_2}\dots\ket{j_N} 
\end{multline}

\noindent  where $ C_{j_1j_2\ldots j_N}$ is the unfactorized coefficient tensor. 

\par The main algorithms driving MPS time--evolution are time--evolving block decimation (TEBD) and the time--dependent variational principle (TDVP). The former applies a series of two--site gates to the MPS at each timestep following a Trotter--Suzuki expansion.  The cost of each gate is dominated by the SVD, which scales as $O( d^3 \chi^3)$ for adjacent tensors in  the basis ordering.  Thus, TEBD  scales as $O(M d^3 \chi^3)$ per step, with $M$  the number of spin coupling terms.  This is much better than full--rank evolution if a reasonably small $\chi$ gives a sufficiently small  error.

Long--range interactions are more costly for TEBD since the target state must be brought together with swap gates.  These also lead to broadly--disseminated entanglement during time evolution.  Since long range interactions entangle distant sites, they also increase the required bond dimension between interstitial sites in the register.  A half--chain entanglement entropy of $S$ between these gives a scaling $\chi \sim O(\exp{S})$, exponentially growing the simulation cost between successive TEBD steps.  These factors limit efficient MPS to systems with one--dimensional entanglement structures.  Our simulations are precisely the opposite: they have many short-- and long--range interactions of comparable magnitude. This gives broadly disseminated entanglement and makes it hard to find an acceptable bond dimension cutoff.  Based on these factors, we anticipate that systems with $N = 32$ will be safely cumbersome for long--timescale TEBD or TDVP simulations.  While other tensor networks can tolerate more complicated entanglement patterns, these are not well--adapted to production use and are likely to remain costly.


\subsection{Total Spin Operator}


\begin{figure}[h]
\begin{center}
{\scriptsize 
\begin{adjustbox}{scale=0.75}
\begin{quantikz}
\qw       & \ctrl{5}                   & \ctrl{6}                  & \ctrl{7}       & \qw & \ldots \hspace{15pt} & \qw & \ctrl{9}    & \qw \\                                        
\lstick[wires=5]{$s$} \qw       & \octrl{5}                   & \octrl{6}                  & \octrl{7}       & \qw & \ldots \hspace{15pt} & \qw & \ctrl{9}    & \qw \\                                        
\qw       & \ocontrol{}                 & \ocontrol{}                & \ocontrol{}     & \qw & \ldots \hspace{15pt} & \qw & \control{}  & \qw \\                                    
          &                             &                            &                 &     &                      &     &             &     \\
\	qw       & \ocontrol{}                 & \ocontrol{}                & \control{}      & \qw & \ldots \hspace{15pt} & \qw & \control{}  & \qw \\                                       
\qw       & \ocontrol{}                 & \control{}                 & \ocontrol{}     & \qw & \ldots \hspace{15pt} & \qw & \control{}  & \qw \\                                         
\lstick[wires=5]{$n$}     \qw           & \gate[wires=1]{S^\alpha}    & \qw                        & \qw             & \qw & \ldots \hspace{15pt} & \qw & \qw         & \qw  \\
\qw       & \qw                         & \gate[wires=1]{S^\alpha}   & \qw             & \qw & \ldots \hspace{15pt}  & \qw & \qw         & \qw \\  
\qw       & \qw                         & \qw                        &  \gate[wires=1]{S^\alpha}             & \qw & \ldots \hspace{15pt}  & \qw & \qw         & \qw \\    
          &                             &                            &                 &     &                      &     &             &  \\
\qw       &  \qw                        & \qw                        & \qw             & \qw & \ldots \hspace{15pt} & \qw &  \gate[wires=1]{S^\alpha}                                                   & \qw
\end{quantikz} 
\end{adjustbox}
}
\end{center}
\caption{{\bf Total spin operator.} 
 Circuit implementing the the total spin operator $S^\alpha_\text{tot}$, where $\alpha \in \{x,y,z, +, -\}$ denotes the constituents. This arrangement requires $s = \lceil \log_2 n\rceil$ ancilla, which must be put into an equal superposition using a Hadamard transformation and uncomputed after calculating the correlation function. }
 \label{fig:total_spin_operator}
\end{figure}

\subsection{Condensed--Matter Models}\label{appendix:condensed_matter_models}

\par To frame our  estimates more broadly, we have calculated the overhead for simulating prototypical   condensed--matter models on a fault--tolerant quantum computer.   These systems carry intrinsic theoretical value, though the classical overhead of large, high--precision calculations lies beyond the scope of existing numerical methods. 

\par The  first model we consider is the  $J_1$--$J_2$ Heisenberg Hamiltonian ($S = 1/2$) on a two--dimensional triangular lattice,

\begin{equation} \label{eq:j1j2_hamiltonian}
H_{12}  =  J_1 \sum_{\substack{\langle i,j\rangle}}  \vec{\sigma}_i \cdot \vec{\sigma}_j +   J_2 \sum_{\substack{\langle\langle i,j\rangle\rangle}} \vec{\sigma}_i \cdot \vec{\sigma}_j,
\end{equation}

\noindent where $\vec{\sigma}_i = (\sigma^x_i,\sigma^y_i,\sigma^z_i)$ is a vector of Pauli operators acting on the $i$--th lattice site. Repeated brackets denote summation over nearest and next--nearest neighbor pairs, respectively. This frustrated spin system is believed to exhibit spin--liquid order at intermediate values of $J_2 / J_1$. Nonetheless, numerical and analytical efforts remain ambiguous regarding the precise nature of this phase \cite{Zhu2015, Hu2015, Sherman2023}.  The $J_1$--$J_2$ model can be encoded in the same manner as our NMR Hamiltonians since both are defined in terms of Pauli strings.

\par Our second reference is the Fermi--Hubbard model, which is  believed to capture the leading behavior of strongly correlated materials.  This problem has garnered significant attention as a target for quantum computation \cite{Babbush2018, Cade2020, Kan2024, Agrawal2024}.	 We consider the single--orbital case on an $N_x \times N_y$  square lattice as specified by the Hamiltonian,

\begin{equation}
H_\text{FH} = -\sum_{\langle i,j \rangle,\alpha} J_{ij} (c_{i,\alpha}^\dagger c_{j,\alpha} + \text{h.c.}) + \sum_i U n_{i,\downarrow} n_{i, \uparrow}.
\end{equation}

\noindent Here $\{c_{i\alpha},c^\dagger_{j\beta}\} = \delta_{ij} \delta_{\alpha \beta}$ are fermionic creation and annihilation operators for a particle of spin $\alpha$ or  $\beta$ on the $i$--th or $j$--th lattice site, respectively.  By extension, we write  $n_{i\alpha} = c^\dagger_{i\alpha} c_{i,\alpha}$ for the fermionic particle number operator.  This system is parameterized by an intersite hopping strength $J_{ij}$ and a many--body interaction $U$ between particles of different spin on the same lattice site.  We  consider an isotropic limit  where $J_{ij} = J$ is constant and the ratio $U/J$ determines the underlying physics.

\par Unlike our other targets, the Fermi--Hubbard model is defined in terms of fermions.  We can tackle this by using the Jordan--Wigner transform to map spinless fermions to qubits,

\begin{eqnarray}\label{eq:fh_hamiltonian}
c_j &=&  (-Z)^{\otimes (j-1)} \otimes \sigma^-_j \otimes I^{\otimes(N-j)}, \\
c^\dagger_j &=& (-Z)^{\otimes (j-1)}   \otimes \sigma^+_j \otimes I^{\otimes(N-j)}, \\
n_j &=& c^\dagger_j c_j = (I_j + Z_j)/2,
\end{eqnarray} 

\noindent where  $\sigma_j^{\pm} = (X_j \pm \imath Y_j)/2$ are spin ladder operators.  To accommodate our spinful case, we double the number of lattice sites within each row to give a $2 N_x \times N_y$ lattice.  We then designate even and odd sublattices as spin--down and spin--up orbitals, respectively, so that row--adjacent pairs of different spin--orbitals can be mapped to the same physical site. Thus, the hopping terms from $H_\text{FH}$ are confined to sublattices of the same parity while the many--body interaction occurs between different spins on the same physical site (which lie on different lattices).  A mapping between the 2d square lattice and a 1d chain is accomplished by defining a zigzag path between rows, giving a constant Pauli string length for coupled sites in a column.  Putting these pieces together, we obtain an amenable representation for our Hamiltonian:

\begin{widetext}
\begin{multline}
H_\text{FH} = \frac{J}{2}\sum_{j=1}^{N_y}\sum_{k=1}^{N_x-1} \Big[X_{(j,2k-1)}  \vec{Z}  X_{(j,2k+1)} + Y_{(j,2k-1)} \vec{Z}  Y_{(j,2k+1)} +  X_{(j,2k)}  \vec{Z}  X_{(j,2k+2)} + Y_{(j,2k)}  \vec{Z} Y_{(j,2k+2)}\Big] + \\
 \frac{ J\, (-1)^{N_x} }{2} \sum_{j=1}^{N_y-1} \sum_{k=1}^{2N_x} \Big[X_{(j,k)} \vec{Z} X_{(j+1,k)} + Y_{(j,k)}\vec{Z} Y_{(j+1,k)} \Big] + \frac{U}{4} \sum_{j=1}^{N_y} \sum_{k=1}^{N_x} (I\ + Z_{(j,2k - 1)})  (I + Z_{(j,2k)})
\end{multline}
\end{widetext}
 
\noindent The parenthetical notation $(j,k) = 2N_x (j-1) + k $ denotes the  index of the subscripted lattice site in the one--dimensional qubit register.  Note that we do not include a chemical potential term since the filling fraction is set by the initial state in this arrangement.  

\par Our resource estimates for the $J_1$--$J_2$ model assume GQSP dynamics generated by Eq.~\ref{eq:j1j2_hamiltonian}, with a $t_\text{max} = 1.0$ s and $J_1 = $ Hz  and $J_2 = 0.5$ Hz.  Similarly, we set $U/J = -4.0$ for Fermi--Hubbard model and adopt a maximal simulation time of  $t_\text{max} = 2 \pi \cdot 200.0$.  This long simulation would set a lower bound on the energy scale for any correlation functions. Hamiltonians are block encoded as an LCU over Pauli strings using the same alias sampling strategy as NMR Hamiltonians. We specify a precision of $0.1\%$ for the Hamiltonian coefficients and a phase angle sequence that reproduces the target unitary with a fidelity of $\epsilon = 0.001$.  It should be noted that more efficient encoding strategies exist for the Fermi--Hubbard model \cite{Babbush2018}.  However, we follow the outlined approach since it is comparable in structure to our other representations.


\subsection{Economic Utility}\label{app_economic_utility}

\par ZULF NMR has the potential to compliment  high--field spectroscopy in some applications and replace it in others \cite{Blanchard2021, Barskiy2024}.  A notable benefit is that spin couplings can be extracted with extremely high precision \cite{Blanchard2013,Wilzewski2017}, including components of the heteronuclear dipolar coupling that are obscured at low field \cite{Blanchard2015}.  These data can be extremely useful when resolving high--resolution molecular structures. The unique nature of ZULF could also permit the immediate determination of molecular chirality, while conventional methods require cumbersome experiments with an auxiliary compound \cite{King2017}. In terms of direct replacement value, a low--cost, benchtop $J$--coupling spectrometer could reduce turnaround for some prosaic tasks in a chemical research setting. Nonetheless,  these tasks require simulation to assign resonance peaks and fit candidate structures to spectra \cite{Wilzewski2017}.  This is a practical bottleneck.

\par With regard to compact instrumentation, low--field techniques based on atomic magnetometry and hyperpolarization can eliminate costly superconducting magnets and cryogenic systems. This leads to lower capital and operational costs, providing a substantial advantage to ZULF technologies. Compact instruments could also have value in field settings, particularly when the goal is to screen an analyte against a database of known molecular fingerprints (e.g., such as environmental monitoring, forensic, and CBRN / defense applications).  This market is currently addressed by portable optical spectroscopies (infrared, Raman) and mass spectrometers, though NMR provides a complimentary and discerning technique.  Small NMR spectrometers could be especially useful when quantifying hazardous materials like explosives or chemical warfare agents, particularly when encountering a new agent.  In this case, the ability to sequester the instrument in a controlled environment could have a notable impact in terms of safety.  While all molecular fingerprinting applications require experimental reference spectra or simulations, the latter is a standout for particularly hazardous materials.

\par There are further advantages for ZULF in laboratory, clinical, and manufacturing contexts. For instance, these spectrometers can be used alongside systems for real--time reaction monitoring \cite{Barskiy2019,Eills2023}, which is difficult with a bulky high--field instrument. ZULF's insensitivity to sample inhomogeneity (e.g., magnetic susceptibility variations) \cite{Blanchard2015,Tayler2018} or conductive environments \cite{,Tayler2019} also permits integration into analytical systems like stopped flow mixers, where metal components or strong electrolytes are likely to be present.  Similar benefits carry over to medical diagnostics or manufacturing processes where NMR spectroscopy  has been difficult to apply. Imaging techniques that observe an entire device --- or provide spatially--resolved chemical shift information --- are another context in which this insensitivity is valuable. An example might be the monitoring of chemical processes in solid--state battery designs.

\par The ZULF setting can also host techniques that are difficult to engineer with conventional NMR.  Notably, the possibility of exploiting robust quantum control might deliver new opportunities for nanoscale measurement and imaging \cite{Jiang2018, Abobeih2019}. Other possibilities include the use of exotic magnetic resonance processes, such as $\beta$-- or $\gamma$--NMR, which have proven valuable as diagnostics for materials science.

\par We can loosely assess the impact of this technology by estimating economic impact for two very specific application classes.


\subsubsection{Small--Molecule Drug Discovery Pipelines}

\par  The most straightforward use case is a small--molecule drug--discovery workflow, though similar considerations hold for molecular design across the broader chemical industry.  This effort could be particularly valuable since roughly 90\% of marketed drugs are small molecules.  We consider the utility of ZULF in replacing high--field spectroscopy for this application within the United States of America.  

\par To quantify value, we assume that a given drug discovery pipeline might handle up to 5000 distinct entities  that would benefit from ZULF NMR in a single year.  While large, this figure estimate accounts for experiments under different environmental conditions, such as when developing production scale--up methods (here we assume a novel use case for ZULF methods). It may also be necessary to acquire multiple ZULF spectra of a given compound to resolve challenging structures.  We assume that computation would be needed to interpret each of these spectra.  For many application it would be necessary to turn over data within 48 hours, though challenging molecules could enjoy a longer timeframe on the order of one to two weeks (e.g., novel natural product leads). A major pharmaceutical company may have 10 robust small--molecule pipelines, and thus the potential for $ 5 \times 10^4$ spectra to  be analyzed annually.  Conversely, a startup may have a single notable pipeline.  

\par The United States has capitalized on 56\% percent of the top 25 pharmaceutical companies, so we will assume that there are 13 major domestic drug manufacturers (albeit with varied degrees of internal research and development) \cite{CBOPharma2021, DDT2024, PharmaExec2023}.  There are also more than 5000 pharmaceutical, biotechnology, and pharmaceutical--supporting enterprises in the US, which are equal or smaller in research volume. We assume that half of these address small--molecule research in some manner.  A good approximation is to treat this contribution as 2500 startup-scale projects with the equivalent of a single robust small--molecule pipeline (even if this is actually adjacent research, e.g., diagnostics, research tool, or adjuvant synthesis).  This amounts to $1.35 \times 10^7$ compounds to be analyzed per year.  The value of ZULF would invariably be justified if the return on a spectrum is roughly $10\%$ of a routine hourly NMR facility rate for high--field NMR (around $\$ 50$ per hour in an academic setting).  This suggests an upper bound of \$65 M in annual value for the pharmaceutical industry.  

\par It is important to note that sufficiently small molecules might not benefit from the use of quantum computation.  In our more complex spectroscopic regimes (e.g., {\bf Heteronuclear + RDCs}), roughly 17\% of the small--molecule entities would be clearly taxing for classical computation.  This could give a lower bound of \$11M in annual value.  The precise details would depend on a more careful quantification of molecular datasets in various industries. 

\subsubsection{Protein Structure Elucidation}

\par Protein NMR is an extremely valuable technique in structural biology.  This method can reveal molecular structures, map their dynamics, and quantify the response to a perturbation (e.g., drug binding) down to the atomic scale.   However, these systems also require numerous time--consuming experiments to obtain a single structure.  We will assume that ZULF  could augment or replace existing NMR techniques for this application.

\par A typical academic grant for protein NMR work would extend over three to five fiscal years and yield five to ten major experimental deliverables / structures.  Using an average over 12 NMR--centric grants from the  National Institutes of Health (USA) in FY24 as a reference, we estimate average direct costs of \$272K per grant in a given year.    Within the budget for this effort, we can assume that between  \$45K and \$100K would be devoted to annual salary.  In an academic setting, these bands  map to (i) a  single graduate student with tuition and benefits and (ii) a postdoc with benefits.  Additional cost overhead is  associated with protein growth, expression, and purification (\$10K to \$50K, including isotopic labeling), other experimental characterization (up to \$10K), and miscellaneous expenses such as travel or publication fees (\$5K to \$10K).  From this, we estimate that \$102K will be used for  NMR instrumentation fees per annum.  If three major structures or datasets of utility are delivered per year, we can estimate that NMR will have a value of \$34K per protein.  This is an upper bound on the value that ZULF methods can deliver. It is important to note that multiple spectra or experiments might be needed to resolve one of these instances. 

\par Our analysis focused an the academic setting, though the financial utility of protein NMR might be even greater in industry. Unfortunately, the lack of open source data makes this difficult to quantify.


\subsection{Physical Resource Estimation}\label{appendix:physical_estimate_methodology}
	
\par Our physical resource estimates are based on  lattice surgery with the rotated surface code \cite{Horsman2012, Litinski2019}.  This is natural for planar devices with   nearest--neighbor  connections, though other hardware layouts can be used with modest overhead.  To guide our analysis, we adopt design principles on parity with Ref.~\cite{Gidney2021}. Doing so maintains a common reference point for overhead though the overall implementation is not highly optimized.

\subsubsection{Fault--Tolerant Layout}

\par A foundational and potentially rate--limiting requirement is the production of magic states. We accomplish this using AutoCCZ factories, which can deliver $\ket{T}$ states by generating of an intermediate $\ket{\text{CCZ}} = \text{CCZ}\ket{+}^{\otimes 3}$   and performing a catalytic  $\ket{\text{CCZ}} \rightarrow 2\ket{T}$ conversion \cite{Gidney2019,Fowler2019}.  In brief, the entry point is an injection of $T$--states at code distance $d_0 = \lfloor d_1 / 2 \rfloor$ using well--established methods \cite{Li2015}.  These  states are  subsequently passed through a round of 15:1 Reed--Muller distillation in `level--1'  $T$--subfactories \cite{Fowler2019}.  The result is a distance $d_1$ refinement that we denote as $\ket{T_1}$.  A number  $\eta$ of the $\ket{T_1}$ states are then fed into a specialized level--2 subfactory that produces a $\ket{\text{CCZ}}$  at distance $d_2 > d_1$.   When desired, catalytic circuitry can mediate the  $\ket{\text{CCZ}} \rightarrow 2\ket{T_2}$ conversion.  Since the level--2 CCZ--factory takes multiple inputs it must be accompanied by
\begin{equation}
\mathcal{N}_{T1} = \left\lceil \eta \cdot \frac{\mathcal{D}_{T1}}{\mathcal{D}_{\text{CCZ}}} \right\rceil
\end{equation}
\noindent of the level--1 $T$--factories to avoid a bottleneck.  Here, $\mathcal{D}_{T1}$ and $\mathcal{D}_{\text{CCZ}}$ are cycle depths for the respective factories while $\eta$ is a constant that depends to the target state ($\eta = 8$ for a terminal $\ket{CCZ}$ and $\eta = 4$ for a terminal $\ket{T}$). The overall cycle depth for a round of magic state distillation is then,
\begin{equation}
\mathcal{D}_\text{distill} = \max (\mathcal{D}_{T1}, \mathcal{D}_{\text{CCZ}} + \mathcal{D}_{\text{cat}}).
\end{equation}

\noindent This expression also includes a $\mathcal{D}_{\text{cat}} = 1\cdot d_2$ cycle overhead for catalysis.  The AutoCCZ factory's spatial footprint is an aggregate of these components,
\begin{equation}
\mathcal{A}_\text{fact} = \eta \, \mathcal{A}_{T1} + \mathcal{A}_\text{CCZ} + \mathcal{A}_\text{cat} + \mathcal{A}_\text{store}.
\end{equation}
\noindent Each contribution $\mathcal{A}_x = 2 n_x d_x^2$ is specified in terms of the number of $n_x$ of logical code tiles in each subcomponent and the number $2d_x^2$ of physical qubits in the corresponding tiles.  The term $\mathcal{A}_\text{store}$ captures routing and storage for level--1 $T$--states (e.g., for caching redundant catalytic states to avoid failure).

\par Our magic state factories aim to deliver $T$--states.  Nonetheless, the $\ket{\text{CCZ}}$ is a valuable magic state since it can directly implement Toffoli gates.  A refined  strategy might leverage both $\ket{\text{CCZ}}$ and $\ket{T_2}$ states, particularly in light of the Toffoli--based logic from our block encoding.   Optimized layouts for QROM access have assumed this execution model  \cite{Lee2021}.   We do not pursue this route since our logical estimates are based on $T$--counts alone.  Instead, we defer to a standard fast block layout which requires only $\mathcal{N}_\text{data} = 2 N_\text{logical} + \sqrt{8 N_\text{logical}} + 1$ tiles for the algorithm, ancilla, and routing \cite{Litinski2019}.  This arrangement permits the consumption of at least one $\ket{T}$ state every reaction cycle.  We assume that magic state factories are  placed around the periphery of this layout.  This is important since our goal is to run a reaction--limited computation.

\begin{figure}[t]
\begin{center}
\setlength\tabcolsep{0pt}
\begin{tabular*}{\linewidth}{@{\extracolsep{\fill}} l |  l  l r  }
\toprule
                 & {\bf Parameter}  & {\bf Symbol}                & {\bf Value }          \\
\midrule\midrule
{\bf Level 0 }   &		            &                             &                       \\
(Injection)      & Volume           &  $\mathcal{V}_{T_1}$        &   100 cells           \\
\midrule
{\bf Level 1 }   &		            &                             &                       \\
($T_1$--Factory) & Area             &  $\mathcal{A}_{T_1}$        &   $4d_1 \times 8d_1$  \\ 
                 & Depth            &  $\mathcal{D}_{T_1}$        &   $5.75 \, d_1$       \\
                 & Volume           &  $\mathcal{V}_{T_1}$        &   1100 cells          \\
\midrule
{\bf Level 2 }   &                  &                             &                       \\
(CCZ, Catalyst)  & Area             &  $\mathcal{A}_\text{CCZ}$   &   $6d_2 \times 3d_2$  \\
                 &                  &  $\mathcal{A}_\text{cat}$   &   $4d_2 \times 4d_2$  \\
                 & Depth            &  $\mathcal{D}_\text{CCZ}$   &   $5.00 \, d_2$       \\
                 &                  &  $\mathcal{D}_\text{cat}$   &   $1.00 \, d_2$       \\
                 & Volume           &  $\mathcal{V}_\text{CCZ}+\mathcal{V}_\text{cat}$   &   1000 cells \\
\midrule
{\bf Hardware}   &                  &                             &                       \\
                 &  Cycle Time      &  $t_\text{cycle}$           &   $1  \, \mu\text{s}$ \\
                 &  Reaction Time   &  $t_\text{react}$           &   $10 \, \mu\text{s}$ \\
\bottomrule 
\end{tabular*}
\end{center}
\caption{ {\bf Parameters for physical resource estimation.}  Input parameters are presented for the AutoCCZ constructions from Ref. \cite{Gidney2019} while minding the improvements from \cite{Gidney2019b,Gidney2021}.}
\label{table:physical_parameters}
\end{figure}

\subsubsection{Error Budget}

\par Our foundational unit is the topological error per spacetime cell,

\begin{equation}
\mathbb{E}(d) = 0.1 \left(\frac{p_\text{phys}}{p_\text{thresh}} \right)^{(d+1)/2},
\end{equation}

\noindent where $d$ is the code distance for the underlying logical qubit, $p_\text{thresh} = 0.01$ is the code error threshold, and $p_\text{phys}$ is the physical error rate (e.g., for a two--qubit gate).  We then subdivide our  total non--algorithmic error,

\begin{equation}\label{eq:ftqc_error_aggregate}
\epsilon_\text{phys} = \mathcal{D}_\text{meas} \mathcal{N}_\text{data}\mathbb{E}(d_2) + \frac{N_T}{2}\epsilon_{T2}.
\end{equation}

\noindent  where  $\mathcal{D}_\text{meas}$ is the measurement depth. The first term captures  topological error for data / routing qubits while the second term accounts for imperfectly prepared magic states.  Following Litinski's execution model, we translate $T$--gates into Pauli product rotations and commute Clifford gates to the end of the circuit \cite{Litinski2019}.  The latter can be absorbed into terminal Pauli product measurements which allows us to forego their accounting in our estimates.  We will assume a worst--case scenario where  $\mathcal{D}_\text{meas}$ is equal to the $T$--depth, though in principle this can be reduced through analysis of the target circuit. Note that \ Eq.~\ref{eq:ftqc_error_aggregate} contains a factor of 1/2 since $\epsilon_{T2}$ corresponds to the catalytic production of two $T$--states.  This error is calculated by propagating error through each stage of distillation,
\begin{eqnarray}
\epsilon_\text{inj}  &=&  \mathcal{V}_\text{inj} \, \mathbb{E}(d_1/2) + p_\text{phys},        \\
\epsilon_\text{T1}   &=&  \mathcal{V}_{T1} \, \mathbb{E}(d_1) + 35\epsilon_\text{inj}^3,      \\
\epsilon_\text{CCZ}  &=&  \mathcal{V}_\text{CCZ} \, \mathbb{E}(d_2) + 28\epsilon_\text{T1}^2, \\
\epsilon_\text{T2}   &=&  \mathcal{V}_\text{cat} \, \mathbb{E}(d_2) + \epsilon_\text{CCZ},   
\end{eqnarray}
\noindent where $\mathcal{V}_n$ is the spacetime volume of the corresponding factory in unit cells (e.g., one distance $d$ logical qubit over $d$ syndrome extraction cycles).  

\subsubsection{Operation and Timing}

Our objective is to run calculations that are limited  by the reaction time $t_\text{react}$ for an error correction cycle.  In this case, the wall runtime is effectively $T_\text{wall} = \mathcal{D}_\text{meas} t_\text{react}$. We will require $\mathcal{N}_\text{fact} = N_T \mathcal{D}_\text{fact} t_\text{cycle}/T_\text{wall}$ magic state factories to maintain this rate by ensuring a consistent pool of $T$--states.  This defines the physical qubit footprint
\begin{equation}
N_\text{phys} = \mathcal{N}_\text{fact}\mathcal{A}_\text{fact} + 2\mathcal{N}_\text{data} d_2^2,
\end{equation}

\noindent for our computation. This quantity is specified  by the minimal code distance that performs our calculation with error $\epsilon_\text{phys} < \epsilon_\text{target}$ according to Eq.~\ref{eq:ftqc_error_aggregate}.

\end{document}